\documentclass[twocolumn,aps,showpacs,superscriptaddress,longbibliography]{revtex4-1}
\usepackage{graphicx}
\usepackage{latexsym}

\begin{document}

\title{Noise reduction in photon counting by exploiting spatial correlations}

\author{Jan Pe\v{r}ina Jr.}
\email{jan.perina.jr@upol.cz} \affiliation{RCPTM, Joint Laboratory
of Optics of Palack\'{y} University and Institute of Physics of
the Czech Academy of Sciences, Faculty of Science, Palack\'{y}
University, 17. listopadu 12, 77146 Olomouc, Czech Republic}

\author{V\'{a}clav Mich\'{a}lek}
\affiliation{Institute of Physics of the Czech Academy of
Sciences, Joint Laboratory of Optics of Palack\'{y} University and
Institute of Physics of CAS, 17. listopadu 50a, 772 07 Olomouc,
Czech Republic}

\author{Ond\v{r}ej Haderka}
\affiliation{Institute of Physics of the Czech Academy of
Sciences, Joint Laboratory of Optics of Palack\'{y} University and
Institute of Physics of CAS, 17. listopadu 50a, 772 07 Olomouc,
Czech Republic}

\begin{abstract}
Joint photocount distributions of a weak twin beam acquired by an
iCCD camera are analyzed with respect to the beam spatial
correlations. A method for extracting these correlations from the
experimental joint photocount distributions is suggested using a
suitable statistical model that quantifies the contribution of
spatial correlations to the joint photocount distributions. In
detail, the profile of twin-beam intensity spatial
cross-correlation function is revealed from the curve that gives
the genuine mean photon-pair number (both photons from a pair are
detected) as a function of the extent of the detection area. Also,
the principle of reducing the noise in photon-number-resolving
detection by using spatial correlations is experimentally
demonstrated.

\end{abstract}

% 42.65.Lm    Parametric down conversion and production of entangled photons
% 42.50.Ar    Photon statistics and coherence theory

\maketitle

\section{Introduction}

Experimental determination of photon-number distributions of weak
optical fields underwent fast development in the last ten years.
Several types of photon-number-resolving detectors including
super-conducting bolometers \cite{Divochiy2008,Lolli2011},
semiconductor arrays of avalanche photodiodes \cite{Afek2009} or
hybrid photomultipliers \cite{Allevi2010} have been developed and
successfully tested under real experimental conditions. Also
intensified CCD (iCCD) cameras with photocathodes composed of a
large number of single-photon sensitive pixels have been
recognized as efficient and practical photon-number-resolving
detectors
\cite{Haderka2005a,Tasca2013,Fickler2013,Chrapkiewicz2014,Chrapkiewicz2014a,Qi2016}.
Among others, these detectors have been applied for the
investigation of photon-number correlations in weak twin beams
containing up to hundreds of photon pairs per pulse
\cite{Allevi2013,PerinaJr2012} as well as to the measurement of
spatial correlations in the fields composed of many photon pairs
\cite{Jost1998,Hamar2010,Tasca2013,Fickler2013,Fickler2014,Jachura2015}.
We note that back-illuminated CCD cameras
\cite{Mosset2004,Lantz2004,Jedrkiewicz2004} as well as
electron-multiplying CCD (EMCCD) cameras \cite{Avella2016}
represent an alternative to iCCD cameras in these experiments:
They have much higher detection efficiencies but also much higher
levels of noise.

As final detection efficiencies and noises of all types of
photon-number-resolving detectors developed up to now cannot be
omitted, the reconstruction of photon-number distributions based
on the experimentally detected photocount histograms represents a
critical step in the characterization of weak optical fields.
Different methods for this reconstruction have been applied,
beginning from a simple linear inversion \cite{Achilles2004} and
ending with the iterative maximum-likelihood approach
\cite{Dempster1977,Rehacek2003}. Also, direct fitting of the
experimental photocount histograms with an anticipated form of the
photon-number distribution has been applied \cite{PerinaJr2013a}.

The experimental characterization of optical fields with a more
complex structure, e.g. weak twin beams, deserves special
attention \cite{Pittman1995,Hamar2010}. Such fields may contain
correlations among their constituting parts not only in photon
numbers, but also in other degrees of freedom. These additional
correlations may influence the photocount histograms observed by
photon-number-resolving detectors. Such histograms may then be in
principle used to reveal these spatial correlations. Twin beams
composed of two entangled (signal and idler) beams represent a
typical example with tight spatial correlations
\cite{Fedorov2008,Joobeur1994,Joobeur1996,Brambilla2004,Brambilla2010,Blanchet2008}.
These correlations may have different forms
\cite{Walborn2010,Molina-Terriza2005} and even spatially
anti-bunched two-photon fields have been generated
\cite{Nogueira2001,Caetano2003}. We note that spatial correlations
of paired fields also enable quantum imaging
\cite{Pittman1995,Gatti2008,Brida2009b,Tasca2013} as well as
sub-shot-noise imaging \cite{Brida2010a} based on sub-shot-noise
correlations of twin beams
\cite{Jedrkiewicz2004,Bondani2007,Blanchet2008,Brida2009a}.
Spatial correlations between the signal and idler photons are
inscribed into the measured data provided that
photon-number-resolving detectors with spatial resolution are used
(e.g., inscribed into the captured frames of an iCCD camera
\cite{Haderka2005,Hamar2010}). This means that the measured data
contain both the information about photon-number statistics and
that about spatial correlations. Simultaneous analysis of the
measured data with respect to both properties is then the most
appropriate and efficient from the point of view of reducing the
noise in the experimental data. On one side and as shown in detail
here, this approach allows to reduce the level of noise in the
determination of intensity spatial cross-correlation functions
with respect to the commonly used method \cite{Haderka2005}. On
the other side and as principally shown by the obtained
experimental curves, spatial correlations allow for partial
reduction of the noise found in the process of photon-pair
counting.

Here, we provide a comprehensive analysis of weak twin beams whose
joint photon-number distribution \cite{Perina1991} and spatial
correlations are simultaneously monitored by an iCCD camera. The
proposed analysis is based upon pairing of the signal and idler
photocounts (counts) identified in the measured camera frames
(pictures) and as such reflecting the detection process (quantum
detection efficiency, noise, spatial properties). The efficiency
of 'software' pairing depends upon the extent of the considered
circular detection area (with the varying radius) that is drawn in
an idler-field detection area (strip) around a point that
corresponds to a given detected signal photon (count) in the
signal-field detection area (strip) [see the drawing in
Fig.~\ref{fig1}]. With the varying radius of the detection area,
the influence of spatial correlations to the obtained joint
signal-idler photocount histograms changes. The larger the
detection area, the larger the number of identified photocount
pairs. However, also the larger the detection area, the larger the
number of identified paired photocounts created randomly from two
counts belonging to different photon pairs (with only one photon
registered) and/or noise photons. To understand qualitatively as
well as quantitatively the relationship between the spatial
correlations and joint photocount distributions, we have developed
a suitable statistical model. Based on this model we have
elaborated a method for revealing the profile of intensity spatial
cross-correlation function from the obtained joint signal-idler
photocount histograms. Contrary to the usual approach for
determining intensity cross-correlation functions that relies on a
homogeneous plateau caused by random photocount pairs, the
developed method makes an estimate of the mean number of random
photocount pairs that is used for eliminating their role in the
obtained experimental data. This more sophisticated approach does
not require huge amount of experimental data (to reach the
homogeneous plateau) and it treats the noise in a more elaborated
(detailed) way when eliminating it from the experimental data.
Also, the pairing procedure identifies (as a by-product) unpaired
counts in the signal and idler strips that originate either in
detection of a noisy photon or just one photon from a photon pair.
Their omission when constructing the joint photocount histograms
results in the noise reduction accompanied by an effective
increase of the detection efficiency of collecting photon pairs
from a twin beam. Intensity of these effects varies with the
extent of the detection area which may be used to efficiently
reduce the noise in photon-pair counting.

To practically demonstrate the approach, we have performed the
analysis of spatially-resolved joint photocount histograms
characterizing a twin beam containing around ten photon pairs on
average and captured by a photocathode of an iCCD camera. The
obtained detection efficiency and the profile of intensity spatial
cross-correlation function have been compared with those found by
the method of absolute detector calibration \cite{PerinaJr2012}
and direct analysis of intensity cross-correlation profiles
\cite{Haderka2005}.

The paper is organized as follows. A general statistical model
appropriate for the detected joint spatially-resolved photocount
histograms is presented in Sec.~II. The approach for revealing
spatial correlations from the joint photocount histograms is
described in Sec.~III. The analysis of experimental data and
comparison with the theoretical model are contained in Sec.~IV.
Sec.~V brings conclusions.

\section{Statistical model for photocount distributions involving
spatial correlations}

In the suggested model, we assume that the joint signal-idler
photon-number distribution $ p $ characterizing a twin beam in
front of the camera can be rewritten as a two-fold convolution of
three photon-number distributions $ p_p $, $ p_s $ and $ p_i $
describing in turn the paired, noise signal and noise idler
components of the overall twin-beam field \cite{PerinaJr2013a}:
\begin{equation}  % 1
 p(n_s,n_i) = \sum_{n=0}^{{\rm min}(n_s,n_i)} p_s(n_s-n)
  p_i(n_i-n) p_p(n).
\label{1}
\end{equation}
In the process of spontaneous parametric down-conversion, the
components are usually assumed in the form of the Mandel-Rice
distribution \cite{Perina1991} defined for a given number $ M_a $
of equally populated modes with mean photon (-pair) number $ B_a $
per mode:
\begin{equation}  % 2
  p_a(n;M_a,B_a) = \frac{\Gamma(n+M_a) }{n!\, \Gamma(M_a)}
  \frac{B_a^n}{(1+B_a)^{n+M_a}}, \hspace{5mm} a={\rm s,i,p}
\label{2}
\end{equation}
and $ \Gamma $ denotes the $ \Gamma $-function.

We derive the corresponding joint signal-idler photocount
distributions in three subsequent steps described in the following
subsections. First, we determine the photocount statistics of
genuine paired counts (caused by photon pairs with both photons
detected), single counts in the signal detection area and single
counts in the idler detection area. Then, in the second step we
determine the distributions of random paired counts, i.e. pairs of
counts found within the corresponding (and varying) signal and
idler detection areas and originating in two different photon
pairs or individual noise counts. Finally, we combine all
contributions together in the third step to arrive at the
appropriate joint signal-idler photocount distributions.

\subsection{Photocount distributions of three components}

We assume that the signal and idler fields illuminate their
detection areas on the photocathode of an iCCD camera
homogeneously. In this case, only the paired component composed of
spatially correlated photon pairs creates spatial correlations.
Their influence to photocount distributions is described as
follows.

\begin{figure}    % fig. 1
 \centerline{\resizebox{0.8\hsize}{!}{\includegraphics{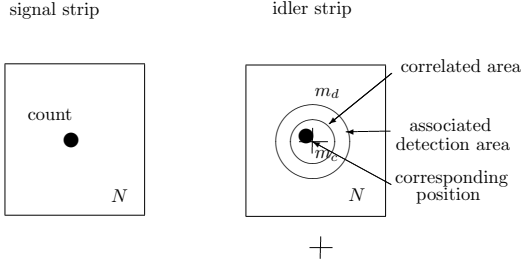}}}
 \vspace{2mm}
 \caption{A typical frame with the signal and idler strips captured by the photocathode:
  Signal (idler) photons are monitored
  inside the signal (idler) detection fields (strips) both containing $ N $ pixels.
  Correlated area of a photon pair extends over $ m_{\rm c} $ pixels.
  The considered detection area with the varying radius covers $
  m_{\rm d} $ pixels. It can be smaller or greater than the correlated
  area.}
\label{fig1}
\end{figure}

Genuine spatial correlations in the detected counts are created
only by photon pairs with both photons detected. For given signal-
($ \eta_{\rm s} $) and idler- ($ \eta_{\rm i} $) field detection
efficiencies, $ N $ pixels in both signal- and idler-field
detection areas (strips), $ m_{\rm c} $ pixels covering the
correlated area, and detection areas with $ m_{\rm d} $ pixels
greater than the correlated area [for the scheme, see
Fig.~\ref{fig1}], the distribution $ \tilde{f}_{\rm p} $ of
genuine paired counts is given by the formula
\begin{equation}  % 3
  \tilde{f}_{\rm p}(c_{\rm p}) = \sum_{n_{\rm p}=0}^{\infty}
  T(c_{\rm p},n_{\rm p};\eta_{\rm s}\eta_{\rm i},0,Nm_{\rm c}) p_{\rm p}(n_{\rm p}) ,
\label{3}
\end{equation}
in which the function $ T $ characterizes the detection process in
a camera. It gives the probability of observing $ c_{\rm p} $
[paired] counts caused by a field with $ n_{\rm p} $ photons
[photon pairs]. For an iCCD camera with $ N $ pixels, detection
efficiency $ \eta $ and mean dark count number $ D $ per one
pixel, the function $ T $ is derived in the form
\cite{PerinaJr2012}:
\begin{eqnarray}     % 4
 T(c,n;\eta,D,N) &=& \left( \begin{array}{c} N \cr c \end{array} \right)
  (1-D)^{N} (1-\eta)^{n} (-1)^{c} \nonumber \\
 & & \hspace{-10mm} \times
  \sum_{l=0}^{c}
  \left( \begin{array}{c} c \cr l \end{array} \right) \frac{(-1)^l}{(1-D)^l}
  \left( 1 + \frac{l}{N} \frac{\eta}{1-\eta}
   \right)^{n} .
\label{4}
\end{eqnarray}
In writing Eq.~(\ref{3}), we have assumed a sufficiently low level
of the signal ($ D_{\rm s} $) and idler ($ D_{\rm i} $) mean dark
count number per pixel such that their contribution to the
creation of random paired counts is negligible.

If only one photon from a photon pair is detected, it causes a
single count in the signal or idler detection fields. The function
$ T $ defined in Eq.~(\ref{4}) allows us to express the
corresponding photocount distributions $ \tilde{f}_{\rm ps} $ and
$ \tilde{f}_{\rm pi} $ in the signal and idler detection fields,
respectively, as follows:
\begin{eqnarray}  % 5
  \tilde{f}_{\rm ps}(c_{\rm s}) &=& \sum_{n_{\rm p}=0}^{\infty}
    T(c_{\rm s},n_{\rm p};\eta_{\rm s}(1-\eta_{\rm i}),0,Nm_{\rm c}) p_{\rm p}(n_{\rm p}) , \nonumber \\
  \tilde{f}_{\rm pi}(c_{\rm i}) &=& \sum_{n_{\rm p}=0}^{\infty}
    T(c_{\rm i},n_{\rm p};\eta_{\rm i}(1-\eta_{\rm s}),0,Nm_{\rm c}) p_{\rm p}(n_{\rm p}).
\label{5}
\end{eqnarray}

The distribution $ \tilde{f}_{\rm p} $ of genuine paired counts is
appropriate only provided that the correlated area is fully
covered by the detection area [$ f_{\rm p}^{\rm reg}(c_{\rm p})
\equiv \tilde{f}_{\rm p}(c_{\rm p}) $ and $ \tilde{f}_{\rm si}(0)
= 1 $, $ \tilde{f}_{\rm si}(c) = 0 $ for $ c=1,\ldots $, see
below]. In an experiment, the detection area is gradually reduced
and so, at certain point, it begins to cover only partially the
correlated area. This fact results in breaking some paired counts.
Introducing probability $ \eta_{\rm d} $ that a count from a
genuine paired count lies within the detection area, the
appropriate distribution $ f_{\rm p}^{\rm reg} $ of genuine paired
counts attains the form
\begin{equation}  % 6
  f_{\rm p}^{\rm reg}(c_{\rm p}) = \sum_{c'_{\rm p}=c_{\rm p}}^{N}
  B(c_{\rm p},c'_{\rm p};\eta_{\rm d}) \tilde{f}_{\rm p}(c'_{\rm p})
\label{6}
\end{equation}
and the binomial distribution $ B $ is written as
\begin{equation}     % 7
 B(c,c';\eta) = \left( \begin{array}{c} c' \cr c \end{array} \right)
   \eta^{c}(1-\eta)^{c'-c} .
\label{7}
\end{equation}
The remaining 'single' counts are governed by distribution $
\tilde{f_{\rm si}} $,
\begin{equation}  % 8
 \tilde{f}_{\rm si}(c_{\rm p}) = \sum_{c'_{\rm p}=c_{\rm p}}^{N}
  B(c_{\rm p},c'_{\rm p};1-\eta_{\rm d}) \tilde{f}_{\rm p}(c'_{\rm p}),
\label{8}
\end{equation}
appropriate for both signal and idler detection fields.

The noise signal and idler components also contribute to counts in
their detection fields. Their photocount distributions $
\tilde{f}_a $, $ a= {\rm s,i} $, are derived from the
corresponding photon-number distributions $ p_a $ applying the
function $ T $ given in Eq.~(\ref{4}):
\begin{equation}  % 9
  \tilde{f}_a(c_a) = \sum_{n_a=0}^{\infty}
   T(c_a,n_a;\eta_a,D_a,N) p_a(n_a) , \hspace{5mm} a={\rm s,i}.
\label{9}
\end{equation}
The photocount distributions $ f_{\rm s} $ and $ f_{\rm i} $
composed of all unpaired counts in the signal and idler detection
fields, respectively, are determined via the convolution:
\begin{equation}  % 10
  f_{a}(c_a) = \sum_{c'_a=0}^{c_a}
   \tilde{f}_{{\rm p}a}(c'_a) \tilde{f}_{a}(c_a-c'_a) , \hspace{5mm} a={\rm s,i}.
\label{10}
\end{equation}

The joint signal-idler photocount distribution $ f_{\rm si} $ of
all counts outside the paired detection areas is finally given by
the two-fold convolution of distributions given in Eqs.~(\ref{8})
and (\ref{10}):
\begin{equation}  % 11
  f_{\rm si}(c_{\rm s},c_{\rm i}) = \sum_{c_{\rm p}=0}^{\min(c_{\rm s},c_{\rm i})}
   \tilde{f}_{\rm si}(c_{\rm p}) \tilde{f}_{\rm s}(c_{\rm s}-c_{\rm p}) \tilde{f}_{\rm i}(c_{\rm i}-c_{\rm p}) .
\label{11}
\end{equation}

The dependence of probability $ \eta_{\rm d} $ introduced in
Eq.~(\ref{6}) on the extent of the detection area is in general
derived from the profile of intensity spatial cross-correlation
function $ t $. Quantifying the extent of detection area
(correlated area) in the number $ m_{\rm d} $ ($ m_{\rm c} $) of
pixels and assuming the constant intensity cross-correlation
function inside the correlated area, we have:
\begin{eqnarray}   % 12
 \eta_{\rm d}(m_{\rm d}) &=& m_{\rm d}/m_{\rm c} , \hspace{10mm} m_{\rm d} < m_{\rm c} , \nonumber \\
             &=& 1 , \hspace{10mm} m_{\rm d}\ge m_{\rm c}.
\label{12}
\end{eqnarray}
In a more realistic case of the Gaussian intensity
cross-correlation profile $ t(\Delta x,\Delta y) = \exp[-(\Delta
x^2 + \Delta y^2)/ R^2 ] / [ \pi R^2 ] $ with radius $ R $ and
considering the circular detection area with radius $ r $, we
arrive at the probability function $ \eta_{\rm d} $ in the simple
form ($ m_{\rm c} = \pi R^2 $, $ m_{\rm d} = \pi r^2 $):
\begin{equation}   % 13
  \eta_{\rm d}(m_{\rm d}) = 1 - \exp(-m_{\rm d}/m_{\rm c}) .
\label{13}
\end{equation}
The model can also be applied to one-dimensional geometry (cuts
across a real correlated area). We have for the Gaussian intensity
profile $ t_{x}(\Delta x) = \exp(-\Delta x^2/ X^2) / (\sqrt{\pi}X)
$ with $ X $ quantifying the extent of the one-dimensional
cross-correlation function ($ m_{\rm c} = 2X $, $ m_{\rm d} = 2x
$)
\begin{equation}  % 14
  \eta_{\rm d}(m_{\rm d}) = {\rm erf}(m_{\rm d}/m_{\rm c}) ;
\label{14}
\end{equation}
$ {\rm erf}(x) \equiv 2/\sqrt{\pi} \int_{0}^{x} \exp(-t^2)dt $
denotes the error function.

\subsection{Random pairing of the signal and idler photocounts}

Individual single counts occurring in the signal and idler
detection fields can accidentally be located at the corresponding
positions within the extent of the detection area. This
configuration creates additional, random, paired counts, whose
number increases with the increasing extent of detection area. To
reveal the appropriate accidental paired photocount distribution,
we first determine probability $ P(c_{\rm p},c_{\rm s},c_{\rm
i};m_{\rm d}) $ that $ c_{\rm s} $ randomly positioned signal
counts together with $ c_{\rm i} $ randomly positioned idler
counts give $ c_{\rm p} $ paired counts within the detection area
covering $ m_{\rm d} $ pixels. If there occur $ c_{\rm s} $ counts
in the signal detection field (strip), the detection areas drawn
around each signal count cover on average a certain part of the
signal detection field, which relative extension is quantified by
probability function $ \eta_{\rm f}(c_{\rm s};m_{\rm d}) $. A
count in the idler detection field has probability $ \eta_{\rm
f}(c_{\rm s};m_{\rm d}) $ to fall into the corresponding area and
thus create an accidental paired count. On the other hand, it
falls outside the corresponding area with probability $
1-\eta_{\rm f}(c_{\rm s};m_{\rm d}) $. For an arbitrary number $
c_{\rm i} $ of idler counts, the effect is described by the
binomial distribution $ B $ already introduced in Eq.~(\ref{7}).
If the number $ c_{\rm p} $ of created paired counts should exceed
the number $ c_{\rm s} $ of signal counts, only the number $
c_{\rm s} $ of paired counts is taken into account. The
probability $ P_{\rm b} $ for characterizing this pairing
procedure in its basic variant is expressed as follows:
\begin{eqnarray}  % 15
 P_{\rm b}(c_{\rm p},c_{\rm s},c_{\rm i};m_{\rm d}) &=&  B(c_{\rm p},c_{\rm i};\eta_{\rm f}(c_{\rm s};m_{\rm d})) \nonumber \\
  & & \hspace{5mm} {\rm for }\;\; c_{\rm p} < {\rm min}(c_{\rm s},c_{\rm i}) ; \nonumber \\
 P_{\rm b}(c_{\rm s},c_{\rm s},c_{\rm i};m_{\rm d}) &=&  \sum_{c_{\rm p}=c_{\rm s}}^{c_{\rm i}}
  B(c_{\rm p},c_{\rm i};\eta_{\rm f}(c_{\rm s};m_{\rm d})) \nonumber \\
  & & \hspace{5mm} {\rm for} \;\; c_{\rm s} \le c_{\rm i} ;  \nonumber\\
  P_{\rm b}(c_{\rm p},c_{\rm s},c_{\rm i};m_{\rm d}) &=& 0 \hspace{15mm}
  {\rm otherwise}.
\label{15}
\end{eqnarray}

The function $ \eta_{\rm f}(c_{\rm s};m_{\rm d}) $ giving the
relative area in the signal detection field (strip) covered by $
c_{\rm s} $ counts, each 'occupying' $ m_{\rm d} $ pixels, is
derived as follows. The first count covers the relative area $ s_1
$ equal to $ s \equiv m_{\rm d}/N $. The second count has already
a smaller empty relative area given by $ 1-s_1 $ to enlarge the
area occupied by the first count. The area occupied on average by
an $ i$-th count is expressed in general as:
\begin{eqnarray}   % 16
 & s_0 = 0, \hspace{3mm} s_1\; =\; s, & \nonumber \\
 & s_i = s \left( 1-\sum_{j=1}^{i-1} s_j \right), \hspace{1mm}
  i=2,\ldots &
\label{16}
\end{eqnarray}
The function $ \eta_{\rm f} $ is then obtained in the simple form:
\begin{equation}   % 17
 \eta_{\rm f}(c_{\rm s};m_{\rm d}) = \sum_{i=0}^{c_{\rm s}} s_{i}(m_{\rm d}) .
\label{17}
\end{equation}

The effect of partial overlapping of the detection areas around
different counts in the signal field (strip) occurs also in the
idler field (strip) inside the area corresponding to the detected
signal counts. To reveal the correction function $ \eta_{\rm
g}(c_{\rm i};c_{\rm s}) $ appropriate for $ c_{\rm s}$ detected
signal counts, each surrounded by $ m_{\rm d} $ detection pixels,
we first determine in parallel the relative areas $
\tilde\eta_{\rm ref} $ and $ \tilde\eta $ inside the area
corresponding to $ c_{\rm s} $ signal counts covered by the
detection areas encircling $ c_{\rm i} $ idler counts without and
with the consideration of possible overlapping, respectively. We
consider that one idler count covers in this area on average $
\eta_{\rm f}(c_{\rm s};m_{\rm d})m_{\rm d} $ pixels. As a
consequence, the relative area $ \tilde\eta_{\rm ref} $ without
the inclusion of overlapping is given as
\begin{equation} % 18
 \tilde\eta_{\rm ref}(c_{\rm i};c_{\rm s}) =
  \frac{ {\rm min}(c_{\rm s},c_{\rm i}) }{ c_{\rm s} }.
\label{18}
\end{equation}
The overlap of the idler detection areas is statistically
quantified by the scheme analogous to that described in
Eqs.~(\ref{16}) and (\ref{17}). It leaves us with the formulas:
\begin{eqnarray}   % 19
 & \tilde\eta(c_{\rm i};c_{\rm s}) = \sum\limits_{i=0}^{c_{\rm i}} \tilde s_{i}(c_{\rm
 s}), &
\label{19} \\
 & \tilde s_0 = 0, \hspace{2mm} \tilde s_1 \;=\; 1/c_{\rm s}, \hspace{2mm}
 \tilde s_i = \tilde s_1 \left( 1-\sum_{j=1}^{i-1} \tilde s_j \right), \hspace{1mm}
  i=2,\ldots . & \nonumber
\end{eqnarray}
Using Eqs.~(\ref{18}) and (\ref{19}), the correction function $
\eta_{\rm g}(c_{\rm i};c_{\rm s}) $ is obtained along the formula:
\begin{equation}   % 20
 \eta_{\rm g}(c_{\rm i};c_{\rm s}) = \frac{\tilde\eta(c_{\rm i};c_{\rm
 s})}{ \tilde\eta_{\rm ref}(c_{\rm i};c_{\rm s}) }.
\label{20}
\end{equation}

This additional correction for the overlap of detection areas
inside the idler field (strip) is involved in the refined model of
pairing and it requires the following modification of probability
$ P $ of pairing written originally in Eq.~(\ref{15}):
\begin{eqnarray}  % 21
 P(c_{\rm p},c_{\rm s},c_{\rm i};m_{\rm d}) &=&
  \sum_{c'_{\rm i}=c_{\rm p}}^{c_{\rm i}} B(c_{\rm p},c'_{\rm i};\eta_{\rm g}(c'_{\rm i};c_{\rm s}))\nonumber \\
 & & \hspace{-10mm} \mbox{} \times
    B(c'_{\rm i},c_{\rm i};\eta_{\rm f}(c_{\rm s};m_{\rm d})) \hspace{5mm}
    {\rm for }\;\; c_{\rm p} < {\rm min}(c_{\rm s},c_{\rm i}) ; \nonumber \\
 P(c_{\rm s},c_{\rm s},c_{\rm i};m_{\rm d}) &=&  \sum_{c_{\rm p}=c_{\rm s}}^{c_{\rm i}}
  \sum_{c'_{\rm i}=c_{\rm p}}^{c_{\rm i}} B(c_{\rm p},c'_{\rm i};\eta_{\rm g}(c'_{\rm i};c_{\rm s})) \nonumber \\
 & & \hspace{-10mm} \mbox{} \times B(c'_{\rm i},c_{\rm i};\eta_{\rm f}(c_{\rm s};m_{\rm d}))
   \hspace{5mm} {\rm for} \;\; c_{\rm s} \le c_{\rm i} ;  \nonumber\\
 P(c_{\rm p},c_{\rm s},c_{\rm i};m_{\rm d}) &=& 0 \hspace{15mm} {\rm otherwise}
\label{21}
\end{eqnarray}
and the binomial distribution $ B $ is written in Eq.~(\ref{7}).

The distribution $ f_{\rm p}^{\rm acc} $ of accidental paired
counts is described via the probability $ P $ given in
Eq.~(\ref{15}) for \emph{the basic model} and in Eq.~(\ref{21})
for \emph{the refined model} in the form:
\begin{equation}   % 22
 f_{\rm p}^{\rm acc}(c_{\rm p};m_{\rm d}) = \sum_{c_{\rm s}=c_{\rm p}}^{N} \sum_{c_{\rm i}=c_{\rm p}}^{N}
   P(c_{\rm p},c_{\rm s},c_{\rm i};m_{\rm d}) f_{\rm si}(c_{\rm s},c_{\rm i}) .
\label{22}
\end{equation}
The remaining unpaired single counts in the signal and idler
detection fields (strips) are left with certain additional
correlations introduced by the pairing procedure. These
correlations are characterized by the conditional distributions $
f_{\rm si}^{\rm acc}(d_{\rm s},d_{\rm i};c_{\rm p}) $ of having $
d_{\rm s} $ signal counts together with $ d_{\rm i} $ idler counts
in a frame with $ c_{\rm p} $ accidental paired counts:
\begin{eqnarray}   % 23
 f_{\rm si}^{\rm acc}(d_{\rm s},d_{\rm i};c_{\rm p},m_{\rm d}) &=& \frac{P(c_{\rm p},c_{\rm p}+d_{\rm s},c_{\rm p}+d_{\rm i};m_{\rm d})
  }{ f_{\rm p}^{\rm acc}(c_{\rm p};m_{\rm d}) } \nonumber \\
 & & \mbox{} \times f_{\rm si}(c_{\rm p}+d_{\rm s},c_{\rm p}+d_{\rm i})  .
\label{23}
\end{eqnarray}
The marginal distributions $ f_{\rm s}^{\rm acc} $ and $ f_{\rm
i}^{\rm acc} $ of the unpaired counts in the signal and idler
detection field, respectively, conditioned by identifying $ c_{\rm
p} $ accidental paired counts are expressed as:
\begin{eqnarray}   % 24
 f_{\rm s}^{\rm acc}(d_{\rm s};c_{\rm p},m_{\rm d}) &=& \sum_{d_{\rm i}=0}^{N}f_{\rm si}^{\rm acc}(d_{\rm s},d_{\rm i};c_{\rm p},m_{\rm d}),
  \nonumber \\
 f_{\rm i}^{\rm acc}(d_{\rm i};c_{\rm p},m_{\rm d}) &=& \sum_{d_{\rm s}=0}^{N}f_{\rm si}^{\rm acc}(d_{\rm s},d_{\rm i};c_{\rm p},m_{\rm d}).
\label{24}
\end{eqnarray}
We note that the original distribution $ f_{\rm si}(c_{\rm
s},c_{\rm i}) $ is then expressed in terms of the distributions $
f_{\rm p}^{\rm acc} $ and $ f_{\rm si}^{\rm acc} $ as follows:
\begin{eqnarray}   % 25
 f_{\rm si}(c_{\rm s},c_{\rm i}) &=& \sum_{c_{\rm p}=0}^{\min(c_{\rm s},c_{\rm i})}
  f_{\rm p}^{\rm acc}(c_{\rm p};m_{\rm d}) \nonumber \\
 & & \mbox{} \times f_{\rm si}^{\rm acc}(c_{\rm s}-c_{\rm p},c_{\rm i}-c_{\rm p};c_{\rm p},m_{\rm d}).
\label{25}
\end{eqnarray}

\subsection{Overall joint signal-idler photocount distributions}

The formula (\ref{25}), when combined with formula (\ref{6}) for
the distribution $ f_{\rm p}^{\rm reg} $ of genuine paired counts,
allows us to determine the joint signal-idler photocount
distribution $ F(c_{\rm s},c_{\rm i}) $ of having $ c_{\rm s} $
counts in the signal field (strip) and $ c_{\rm i} $ counts in the
idler field (strip):
\begin{eqnarray}   % 26
 F(c_{\rm s},c_{\rm i}) &=& \sum_{c_{\rm p}=0}^{\min(c_{\rm s},c_{\rm i})} \sum_{c'_{\rm p}=0}^{c_{\rm p}}
  f_{\rm p}^{\rm reg}(c_{\rm p}-c'_{\rm p}) f_{\rm p}^{\rm acc}(c'_{\rm p};m_{\rm d}) \nonumber \\
 & & \mbox{} \times f_{\rm si}^{\rm acc}(c_{\rm s}-c_{\rm p},c_{\rm i}-c_{\rm p};c'_{\rm p},m_{\rm d}).
\label{26}
\end{eqnarray}
Moreover, the distribution $ F_{\rm p} $ of paired counts, both
genuine and accidental, is given as:
\begin{equation}   % 27
 F_{\rm p}(c_{\rm p};m_{\rm d}) = \sum_{c'_{\rm p}=0}^{c_{\rm p}}
  f_{\rm p}^{\rm reg}(c_{\rm p}-c'_{\rm p}) f_{\rm p}^{\rm acc}(c'_{\rm p};m_{\rm d}).
\label{27}
\end{equation}

The joint distribution $ F_{\rm si} $ of unpaired signal and idler
counts is expressed along the relation:
\begin{equation}   % 28
 F_{\rm si}(d_{\rm s},d_{\rm i};m_{\rm d}) = \sum_{c_{\rm p}=0}^{N}
  f_{\rm p}^{\rm acc}(c_{\rm p};m_{\rm d}) f_{\rm si}^{\rm acc}(d_{\rm s},d_{\rm i};c_{\rm p},m_{\rm d}).
\label{28}
\end{equation}
Similarly, the distribution $ F_a $ of unpaired counts in
detection field $ a $, $ a={\rm s,i} $, is determined by the
simple formula:
\begin{equation}   % 29
 F_a(d_a;m_{\rm d}) = \sum_{c_{\rm p}=0}^{N}
  f_{\rm p}^{\rm acc}(c_{\rm p};m_{\rm d}) f_a^{\rm acc}(d_a;c_{\rm p},m_{\rm d}), \hspace{2mm} a={\rm s,i}.
\label{29}
\end{equation}

The numbers of unpaired counts in both detection fields (strips)
can be reduced by considering only those counts occurring in the
detection areas around the identified paired counts. This means
the reduction of noise in the photon counting that characterizes
the twin beam. The distributions $ F^{\rm red}(c_{\rm s},c_{\rm
i};m_{\rm d}) $ and $ F_{\rm si}^{\rm red}(d_{\rm s},d_{\rm
i};m_{\rm d}) $ appropriate for this case are theoretically
determined by the modified expressions in Eqs.~(\ref{26}) and
(\ref{28}) that involve the binomial distributions $ B $ to
account for the reduction of the number of unpaired counts:
\begin{eqnarray}   % 30 - 31
 F^{\rm red}(c_{\rm s},c_{\rm i};m_{\rm d}) &=& \sum_{c_{\rm p}=0}^{\min(c_{\rm s},c_{\rm i})} \sum_{c'_{\rm p}=0}^{c_{\rm p}}
  f_{\rm p}^{\rm reg}(c_{\rm p}-c'_{\rm p}) f_{\rm p}^{\rm acc}(c'_{\rm p};m_{\rm d}) \nonumber \\
 & & \hspace{-20mm} \mbox{} \times \sum_{d'_{\rm s}=c_{\rm s}-c_{\rm p}}^{N} \sum_{d'_{\rm i}=c_{\rm i}-c_{\rm p}}^{N}
  B(c_{\rm s}-c_{\rm p},d'_{\rm s},\eta_{\rm f}(c_{\rm p};m_{\rm d})) \nonumber \\
 & & \hspace{-20mm} \mbox{} \times B(c_{\rm i}-c_{\rm p},d'_{\rm i},\eta_{\rm f}(c_{\rm p};m_{\rm d}))
  f_{\rm si}^{\rm acc}(d'_{\rm s},d'_{\rm i};c'_{\rm p},m_{\rm d}),
\label{30} \\
 F_{\rm si}^{\rm red}(d_{\rm s},d_{\rm i};m_{\rm d}) &=& \sum_{c_{\rm p}=0}^{\min(c_{\rm s},c_{\rm i})} \sum_{c'_{\rm p}=0}^{c_{\rm p}}
  f_{\rm p}^{\rm reg}(c_{\rm p}-c'_{\rm p}) f_{\rm p}^{\rm acc}(c'_{\rm p};m_{\rm d}) \nonumber \\
 & & \hspace{-20mm} \mbox{} \times \sum_{d'_{\rm s}=d_{\rm s}}^{N} \sum_{d'_{\rm i}=d_{\rm i}}^{N}
  B(d_{\rm s},d'_{\rm s},\eta_{\rm f}(c_{\rm p};m_{\rm d})) \nonumber \\
 & & \hspace{-20mm} \mbox{} \times B(d_{\rm i},d'_{\rm i},\eta_{\rm f}(c_{\rm p};m_{\rm d}))
  f_{\rm si}^{\rm acc}(d'_{\rm s},d'_{\rm i};c'_{\rm p},m_{\rm
  d}).
\label{31}
\end{eqnarray}
The distributions $ F_a^{\rm red}(d_a;m_{\rm d}) $ of unpaired
counts in field $ a $, $ a={\rm s,i} $, are easily derived from
the distribution $ F_{\rm si}^{\rm red} $ given in Eq.~(\ref{31})
as follows:
\begin{eqnarray}   % 32
 F_s^{\rm red}(d_s;m_{\rm d}) &=& \sum_{d_{\rm i}=0}^{N}
  F_{\rm si}^{\rm red}(d_{\rm s},d_{\rm i};m_{\rm d}) ,
  \nonumber \\
 F_i^{\rm red}(d_i;m_{\rm d}) &=& \sum_{d_{\rm s}=0}^{N}
  F_{\rm si}^{\rm red}(d_{\rm s},d_{\rm i};m_{\rm d}) .
\label{32}
\end{eqnarray}

Photocount moments of different orders derived from the above
distributions represent important characteristics of the detected
fields. Their combinations allow, among others, the determination
of covariance $ C $ of fluctuations of photocount numbers as well
as sub-shot-noise parameter $ R $, both quantifying mutual
correlations between the numbers of photocounts in two detection
fields:
\begin{eqnarray}   % 33-34
 C &=& \frac{ \langle \Delta c_{\rm s} \Delta c_{\rm i} \rangle}{
  \sqrt{\langle (\Delta c_{\rm s})^2 \rangle \langle (\Delta c_{\rm i})^2 \rangle} },
\label{33} \\
 R &=& \frac{ \langle [\Delta(c_{\rm s}-c_{\rm i})]^2\rangle}{ \langle c_{\rm s}
  \rangle + \langle c_{\rm i} \rangle } .
\label{34}
\end{eqnarray}

\section{Determination of intensity spatial cross-correlation functions}

The profile $ t $ of intensity spatial cross-correlation function
is imprinted into the dependence of the mean number $ \langle
c_{\rm p} \rangle^{\rm reg} $ of genuine paired counts on the
number $ m_{\rm d} $ of detection pixels. This curve can be
approximately derived from the experimental data that provide the
mean number $ \langle c_{\rm p} \rangle $ of all identified paired
counts once we estimate the mean number $ \langle c_{\rm p}
\rangle^{\rm acc,exp} $ of accidental experimental paired counts.
This estimated distribution $ f_{\rm p}^{\rm acc,exp} $ is given
by the modified Eq.~(\ref{22}),
\begin{eqnarray}   % 35
 f_{\rm p}^{\rm acc,exp}(c_{\rm p},m_{\rm d}) &=& \sum_{d_{\rm s}=c_{\rm p}}^{N} \sum_{d_{\rm i}=c_{\rm p}}^{N}
   P(c_{\rm p},d_{\rm s},d_{\rm i};m_{\rm d}) \nonumber \\
  & & \mbox{} \times F_{\rm si}^{\rm exp}(d_{\rm s},d_{\rm i};m_{\rm d}^0),
\label{35}
\end{eqnarray}
in which the distribution $ F_{\rm si}^{\rm exp} $ characterizes
the experimental unpaired counts observed for certain suitable
number $ m_{\rm d}^0 $ of detection pixels. The analysis of
covariance $ C_{\Delta d} $ of the joint signal-idler distribution
$ F_{\rm si}^{\rm exp}(d_{\rm s},d_{\rm i}) $ of unpaired counts
considered as a function of the number $ m_{\rm d} $ of detection
pixels suggests the natural choice implicitly expressed as $
C_{\Delta d}(m_{\rm d}^0) = 0 $. The reason is that the covariance
$ C_{\Delta d} $ is positive for $ m_{\rm d} < m_{\rm c} $ as it
includes genuine paired counts that did not fit into the too small
detection area. On the other hand, for $ m_{\rm d}
> m_{\rm c} $ the pairing procedure 'generates' accidental paired counts
with correlations that have to be compensated in the distribution
$ F_{\rm si}^{\rm exp} $ of the remaining unpaired counts. In more
detail, for $ m_{\rm d} > m_{\rm c} $ the distribution $ f_{\rm
si} $ in Eq.~(\ref{25}) exhibits no correlations between counts $
c_{\rm s} $ and $ c_{\rm i} $. Subsequent application of the
pairing procedure of Eq.~(\ref{22}) removes some paired counts
'outside' the distribution $ f_{\rm si} $ which introduces
negative correlations into the resultant distribution $ F_{\rm si}
$ given in Eq.~(\ref{28}). These correlations result in negative
values of the covariance $ C_{\Delta d} $ found for $ m_{\rm d} >
m_{\rm c} $ (see Fig.~\ref{fig6} below).

The looked-for 'spatially integrated' profile $ \tilde{t} $ of
intensity cross-correlation function $ t $ drawn as a function of
the number $ m_{\rm d} $ of detection pixels can approximately be
inferred from the normalized experimental mean number $ \langle
c_{\rm p} \rangle^{\rm reg,exp}_{\rm n} $ of genuine paired counts
$ \bigl( \langle c_{\rm p} \rangle^{\rm reg,exp}_{\rm n}(m_{\rm
d}) \equiv [ \langle c_{\rm p} \rangle (m_{\rm d}) - \langle
c_{\rm p} \rangle^{\rm acc,exp}(m_{\rm d})] / [ \langle c_{\rm p}
\rangle (N) - \langle c_{\rm p} \rangle^{\rm acc,exp}(N)] \bigr) $
using the following formula:
\begin{equation}  % 36
 \langle c_{\rm p} \rangle^{\rm reg,exp}_{\rm n}(m_{\rm d}) = \int_{0}^{m_{\rm d}}
 d m'_{\rm d}\, \tilde{t}^{\rm app}(m'_{\rm d}).
\label{36}
\end{equation}
Assuming as an example a rotationally invariant correlated area
with profile $ t $ described above Eq.~(\ref{13}), the inversion
of Eq.~(\ref{36}) leaves us with the formula
\begin{equation}  % 37
 \tilde{t}^{\rm app}_R(m_{\rm d}) =  \frac{d \langle c_{\rm p} \rangle^{\rm
  reg,exp}_{\rm n}(m_{\rm d}) }{ dm_{\rm d}}.
\label{37}
\end{equation}
When analyzing photocount correlations in rectangular detection
areas parallel, e.g., to the $ x $ axis, we immediately reveal the
profile $ t_x^{\rm app} $ of the intensity spatial
cross-correlation function in this direction:
\begin{equation}  % 38
 t_x^{\rm app}(m_{\rm d}) = \frac{d \langle c_{\rm p} \rangle^{\rm
  reg,exp}_{\rm n}(m_{\rm d}) }{ dm_{\rm d}}.
\label{38}
\end{equation}

\section{Experimental results and their interpretation}

\begin{figure}         % fig 2
 \centerline{\resizebox{0.8\hsize}{!}{\includegraphics{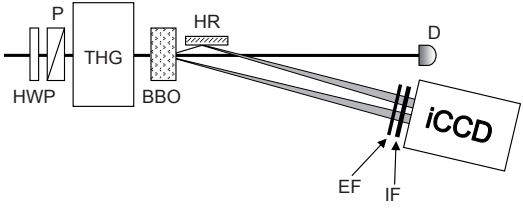}}}
 \vspace{2mm}
 \caption{Scheme of the experiment. A Ti:sapphire laser beam is
  transformed into its third harmonics (THG, 280 nm) that pumps a BaB$ {}_2 $O$ {}_4 $
  (BBO) nonlinear crystal. Nearly degenerate signal and idler (steered by high-reflectivity mirror HR)
  beams are selected using 14-nm-wide bandpass filter IF and detected
  in two detection fields (strips) on a photocathode of iCCD camera. Long-pass (above 490 nm) filter EF diminishes the noise.
  Intensity of the pump beam monitored by detector D is actively stabilized (rms below 0.3\%) using
  motorized half-wave plate HWP followed by polarizer P.}
\label{fig2}
\end{figure}

The experiment was performed with a twin beam centered at the
wavelength 560~nm and originating in a non-collinear type-I
interaction in a 5-mm long BaB$ {}_2 $O$ {}_4 $ crystal pumped by
the third harmonics of a femtosecond cavity dumped Ti:sapphire
laser (pulse duration 150~fs, central wavelength 840~nm)
\cite{Hamar2010}. An iCCD camera Andor DH334-18U-63 was used to
capture individual photocounts in two detection strips, one for
the signal field, the other for the idler field (see
Fig.~\ref{fig2}). Both detection fields were covered by $ N=6500 $
pixels and suffered from $ D = 0.2/N $ mean dark counts per pixel.
The experiment was repeated $ 1.2 \times 10^6 $ times. Before
applying the developed model to the experimental data, we
performed the standard analysis of the joint signal-idler
photocount histogram \cite{PerinaJr2013a} that provided both
parameters of the twin beam ($ B_{\rm p}=0.032 $, $ M_{\rm p} =
280 $, $ B_{\rm s}= 8.2$, $ M_{\rm s} =0.009 $, $ B_{\rm i}= 4.7
$, and $ M_{\rm i}= 0.033 $, relative errors of the parameters are
better than 7\%) and the signal- and idler-field detection
efficiencies ($ \eta_{\rm s} = 0.228 \pm 0.005 $, $ \eta_{\rm i} =
0.223 \pm 0.005 $). We note that the photon-number distributions $
p_a(n) $, $ a={\rm s,i} $, of the noise fields with numbers $ M_a
$ of modes considerably lower than 1 are sharply localized around
$ n=0 $ which is a consequence of the specific electronic response
of the iCCD camera. Values of the determined parameters are used
to derive the predictions about the dependence of the observed
quantities on the number $ m_{\rm d} $ of detection pixels. These
predictions are obtained both for \emph{the basic} as well as
\emph{the refined models} of the pairing procedure, as described
in Eqs.~(\ref{15}) and (\ref{21}).

\begin{figure}         % fig. 3a,b
 \centerline{(a)\hspace{3mm}\resizebox{0.7\hsize}{!}{\includegraphics{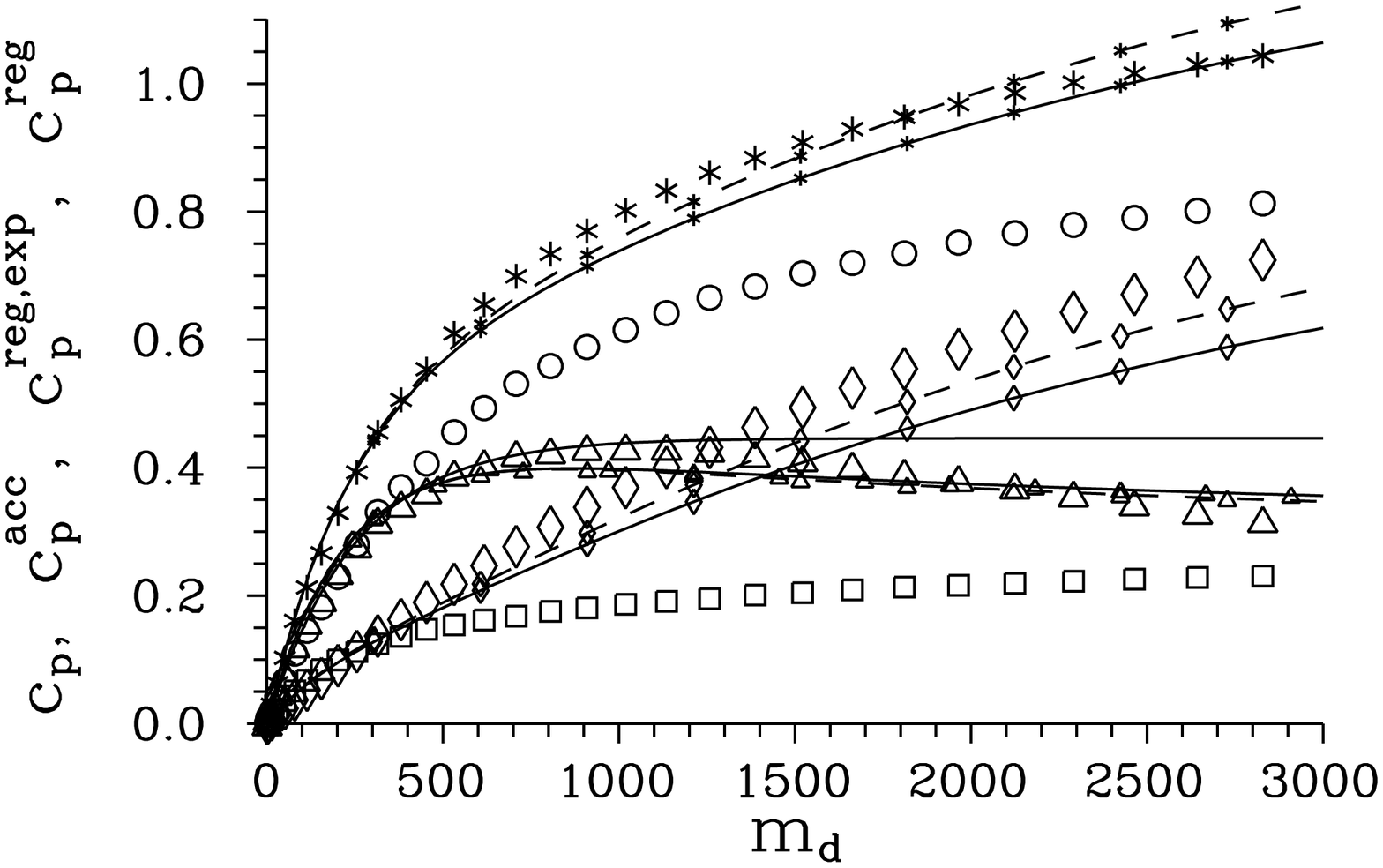}}}
 \vspace{2mm}
 \centerline{(b)\hspace{3mm}\resizebox{0.7\hsize}{!}{\includegraphics{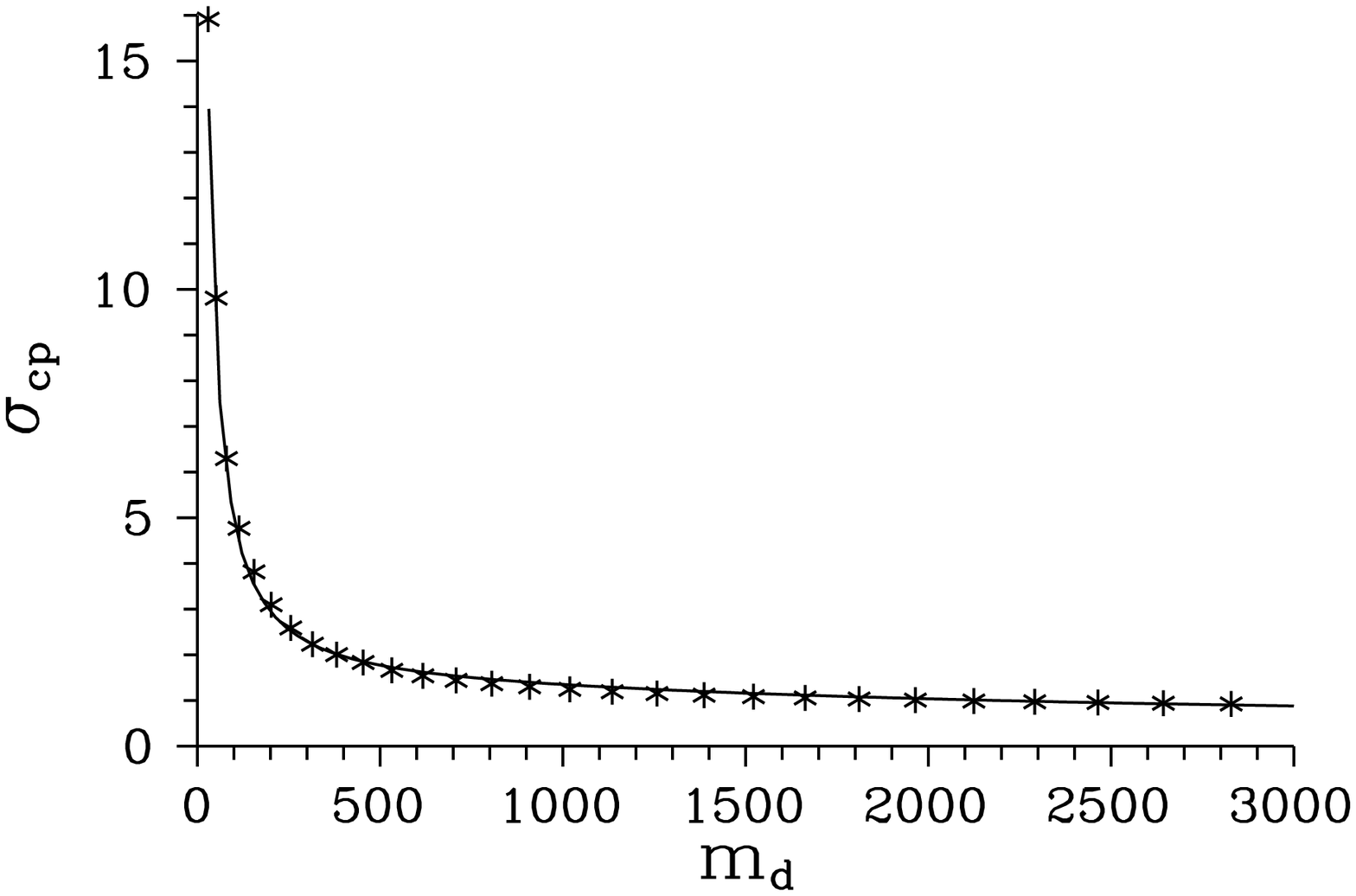}}}
 \vspace{2mm}
 \caption{(a) Mean numbers $ \langle c_{\rm p} \rangle $ ($ \ast $), $ \langle c_{\rm p} \rangle^{\rm acc}
  $ ($ \diamond $), $ \langle c_{\rm p} \rangle^{\rm reg,exp}
  \equiv \langle c_{\rm p} \rangle - \langle c_{\rm p} \rangle^{\rm
  acc} $ ($ \triangle $), and $ \langle c_{\rm p} \rangle^{\rm reg} $ (plain coinciding curves) of in turn all,
  accidental, experimental genuine and theoretical genuine paired counts
  and (b) relative variance $ \sigma_{c_{\rm p}} \equiv ( \langle c_{\rm p}^2 \rangle -
  \langle c_{\rm p} \rangle^2 )/ \langle c_{\rm p} \rangle^2 $ as they depend on the number $ m_{\rm d} $ of detection
  pixels. The mean number $ \langle c_{\rm p} \rangle^{\rm
  reg,exp}$ of experimental genuine paired counts is determined
  using the distribution $ f_{\rm p}^{\rm acc,exp} $ in
  Eq.~(\ref{35}) with $ m_{\rm d}^0 = 290 $. Solid (dashed) curves originate in
  the refined (basic) model, isolated points are derived from the experimental data.
  For comparison, mean numbers $ \langle \tilde{c}_{\rm p} \rangle^{\rm acc}
  $ ($ \Box $) and $ \langle \tilde{c}_{\rm p} \rangle^{\rm
  reg,exp} $ ($ \circ $) obtained from the experimental data
  along Eq.~(\ref{35}) in which the distribution $ F_{\rm si}^{\rm exp} $ is taken
  for the varying number $ m_{\rm d} $ of detection pixels is
  plotted. Relative experimental errors for $ \langle c_{\rm p} \rangle $, $ \langle
  c_{\rm p} \rangle^{\rm acc} $, $ \langle
  \tilde{c}_{\rm p} \rangle^{\rm acc} $, $ \langle c_{\rm p} \rangle^{\rm
  reg,exp} $, $ \langle \tilde{c}_{\rm p} \rangle^{\rm
  reg,exp} $, and $ \sigma_{c_{\rm p}} $ are lower than in turn
  2\%, 1\%, 1\%, 3\%, 3\%, and 2\%.}
 \label{fig3}
\end{figure}

A curve giving the number $ \langle c_{\rm p} \rangle $ of
detected paired counts as a function of the number $ m_{\rm d} $
of detection pixels is the most important curve in the analysis
[see Fig.~\ref{fig3}(a)]. The number $ \langle c_{\rm p} \rangle $
of detected paired counts increases with the increasing number $
m_{\rm d} $ of detection pixels for two reasons. First, the number
of broken genuine paired counts decreases with the increasing
number $ m_{\rm d} $ of detection pixels due to the finite extent
of the correlated area. Second, the number $ \langle c_{\rm p}
\rangle^{\rm acc} $ of accidental paired counts also naturally
increases with $ m_{\rm d} $. As the pairing procedure is stronger
in the basic model compared to the refined one, the numbers $
\langle c_{\rm p} \rangle^{\rm acc} $ of accidental paired counts
as well as the numbers $ \langle c_{\rm p} \rangle $ of all paired
counts predicted by the theory are greater for the basic model
[compare solid and dashed curves with symbols plotted in
Fig.~\ref{fig3}(a)]. The first mechanism that breaks the genuine
paired counts depends on the profile of the correlated area and,
as discussed in the previous section, this profile can be
extracted from the curve giving the number $ \langle c_{\rm p}
\rangle^{\rm reg} $ of genuine paired counts. The application of
Eq.~(\ref{35}) (for details, see below) gives us an experimental
estimate for the number $ \langle c_{\rm p} \rangle^{\rm acc} $ of
accidental paired counts and the number $ \langle c_{\rm p}
\rangle^{\rm reg,exp} $ of experimental genuine paired counts is
then determined simply as $ \langle c_{\rm p} \rangle^{\rm
reg,exp} = \langle c_{\rm p} \rangle - \langle c_{\rm p}
\rangle^{\rm acc} $. The resultant points obtained from the
experimental data as well as the theoretical predictions of both
models are plotted in Fig.~\ref{fig3}(a).

\begin{figure}         % fig. 4
 \centerline{\resizebox{0.9\hsize}{!}{\includegraphics{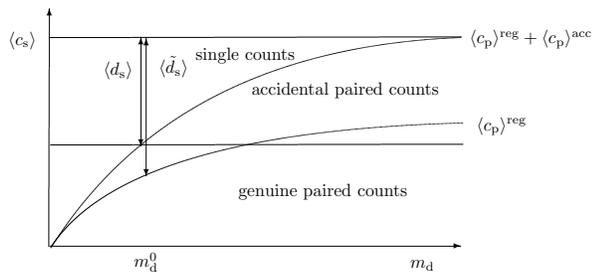}}}
  % \vspace{2mm}
 \caption{Schematic division of counts in the signal detection field (strip) as the number $ m_{\rm d} $
 of detection pixels varies. The constant number $ \langle c_{\rm s} \rangle $ of overall counts is
 composed of mean numbers $ \langle c_{\rm p} \rangle^{\rm reg} $ of genuine paired counts,
 $ \langle c_{\rm p} \rangle^{\rm acc} $ of accidental paired counts and $ \langle d_{\rm s} \rangle $ of single (unpaired) counts.
 Whereas only the number $ \langle d_{\rm s} \rangle $ of unpaired counts is provided directly in the experiment,
 the number $ \langle \tilde{d}_{\rm s} \rangle \equiv \langle d_{\rm s} \rangle + \langle c_{\rm p}\rangle^{\rm acc}
 $ would be needed in the model to estimate the number $ \langle c_{\rm p} \rangle^{\rm acc} $ of accidental paired counts.}
\label{fig4}
\end{figure}

In Fig.~\ref{fig3}(a), the comparison of experimental points and
theoretical curves for $ \langle c_{\rm p} \rangle^{\rm reg,exp} $
with the curve for $ \langle c_{\rm p} \rangle^{\rm reg} $ giving
the actual number of genuine paired counts, that is accessible
only in the model, shows that the experimental numbers $ \langle
c_{\rm p} \rangle^{\rm acc} $ of accidental paired counts are
overestimated. The number $ m_{\rm d}^0 = 290 $ of detection
pixels and the corresponding experimental signal-idler histogram $
F_{\rm si}^{\rm exp}(d_{\rm s},d_{\rm i};m_{\rm d}^0) $ of
unpaired counts have been used in Eq.~(\ref{35}) to arrive at the
numbers $ \langle c_{\rm p} \rangle^{\rm acc} $ of accidental
paired counts plotted in Fig.~\ref{fig3}(a). Decrease of the
number $ \langle c_{\rm p} \rangle^{\rm reg,exp}(m_{\rm d}) $ of
genuine paired counts observed in Fig.~\ref{fig3}(a) for greater
values of $ m_{\rm d} > 3m_{\rm d}^0 $ is apparently caused by
overestimating the number $ \langle c_{\rm p} \rangle^{\rm acc} $
of accidental paired counts. This behavior originates in the fact
that the number $ \langle c_{\rm p} \rangle^{\rm acc} $ of
accidental paired counts is determined via the pairing procedure
described in Eqs.~(\ref{15}) or (\ref{21}) that requires the joint
distribution of single counts in the signal and idler fields
(strips). This distribution does not include only genuine paired
counts and as such it is available only in the model. In the
experimental data the genuine and accidental paired counts cannot
be separated. This fact requires to replace the needed joint
distribution of unpaired counts in the signal and idler detection
fields by a suitable experimental distribution that, however,
cannot incorporate single counts glued into accidental paired
counts. If we consider such experimental distribution as a
function of the number $ m_{\rm d} $ of detection pixels,
underestimation of the number $ \langle c_{\rm p} \rangle^{\rm
acc} $ of accidental paired counts occurs [compare the
experimental symbols $ \circ $ and $ \Diamond $ in
Fig.~\ref{fig3}(a)]. To cope with this effect, we assume in the
pairing procedure given in Eq.~(\ref{35}) the experimental
distribution $ F_{\rm si}^{\rm exp} $ obtained for a certain fixed
value of the number $ m_{\rm d}^0 $ of detected pixels that
roughly covers the correlated area. The chosen value of $ m_{\rm
d}^0 $ determines the behavior of the estimated values of the
number $ \langle c_{\rm p} \rangle^{\rm acc} $ of accidental
paired counts with respect to the true ones. As follows from the
drawing in Fig.~\ref{fig4}, if the chosen value of $ m_{\rm d}^0 $
is too large, the numbers $ \langle c_{\rm p} \rangle^{\rm acc} $
of accidental paired counts are underestimated in the whole range
of $ m_{\rm d} $ because the corresponding experimental
distribution $ F_{\rm si}^{\rm exp} $ describes the signal and
idler fields that are too weak. On the other hand, there occurs
overestimation of the numbers $ \langle c_{\rm p} \rangle^{\rm
acc} $ of accidental paired counts for greater values of $ m_{\rm
d} $ provided that the chosen value of $ m_{\rm d}^0 $ is smaller.
This is our case with $ m_{\rm d}^0 = 290 $. However, such choice
gives us better estimate for the numbers $ \langle c_{\rm p}
\rangle^{\rm acc} $ of accidental paired counts for the numbers $
m_{\rm d}$ of detection pixels comparable to the number $ m_{\rm
c} $ of pixels of the correlated area. As the determination of the
profile of the correlated area is addressed here, we concentrate
our attention to the area with $ m_{\rm d} < 600 $ which justifies
our choice $ m_{\rm d}^0 = 290 $. We note that the numbers $
\langle c_{\rm p} \rangle^{\rm acc} $ of accidental paired counts
for small values of $ m_{\rm d}$ are always underestimated, but
this behavior is not important as the true values of $ \langle
c_{\rm p} \rangle^{\rm acc} $ are small in this case and so they
can be omitted.

With the increasing number $ m_{\rm d} $ of detection pixels the
correlated area with $ m_{\rm c} $ pixels is gradually covered by
the detection area and this fact results in dramatic decrease of
the relative variance $ \sigma_{c_{\rm p}} $ of the number of
detected paired counts, as documented in Fig.~\ref{fig3}(b). The
curve in Fig.~\ref{fig3}(b) clearly indicates appropriate numbers
$ m_{\rm d} $ of detection pixels for which the spatial
correlations are practically lost.

\begin{figure}         % figs. 5a,b
 \centerline{\resizebox{0.46\hsize}{!}{\includegraphics{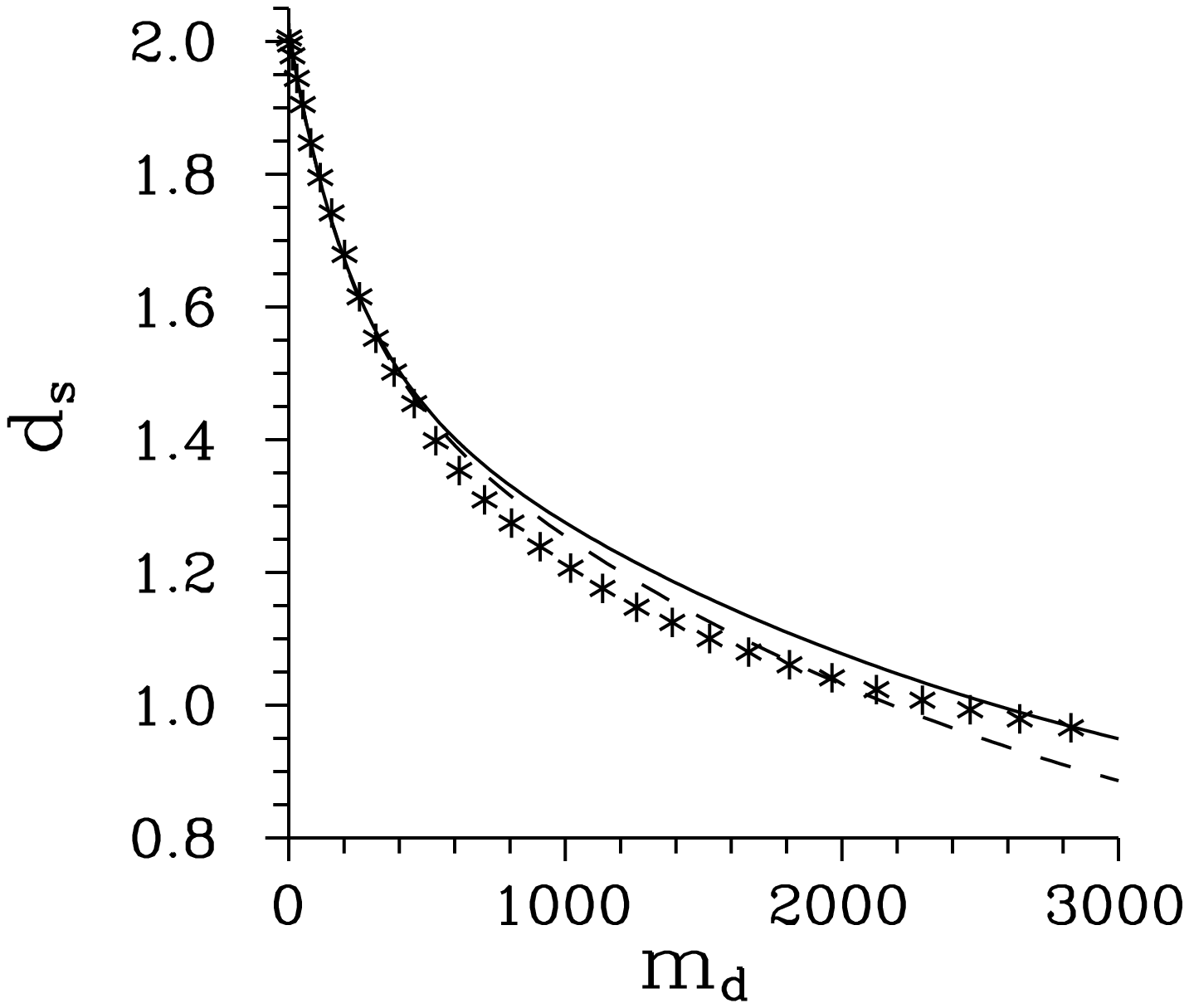}}
  \hspace{3mm}\resizebox{0.46\hsize}{!}{\includegraphics{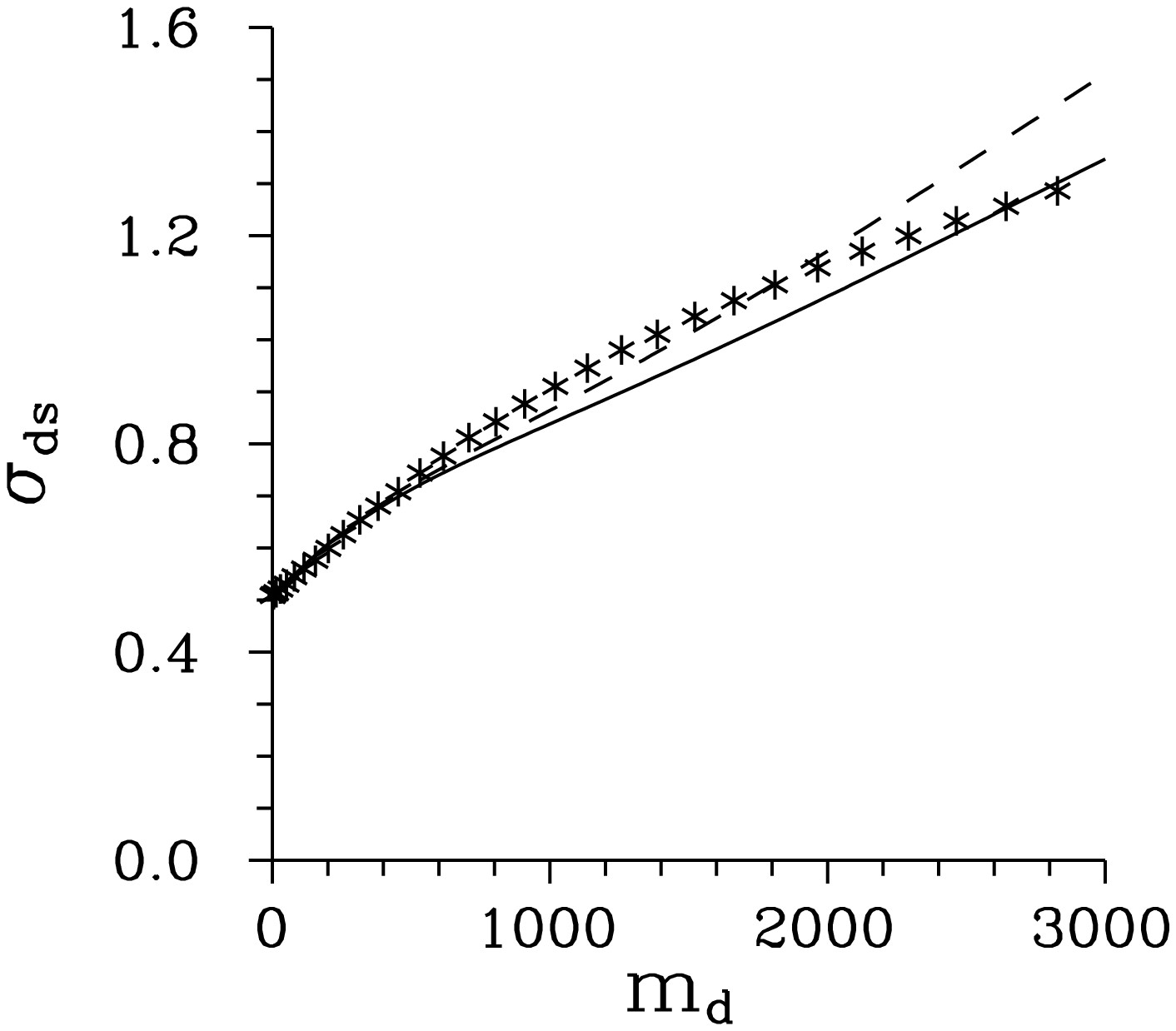}}}
  \centerline{(a) \hspace{.2\textwidth} (b)}
  % \vspace{2mm}
 \caption{(a) Mean number $ \langle d_{\rm s} \rangle $ of signal unpaired counts
  and (b) relative variance $ \sigma_{d_{\rm s}} $ of their distribution as functions of
  the number $ m_{\rm d} $ of detection pixels. Solid (dashed) curves originate in the refined (basic) model, isolated points are
  determined from the experimental data with relative errors lower than 1\%.}
\label{fig5}
\end{figure}

\begin{figure}         % figs. 6a,b
 \centerline{(a)\hspace{3mm}\resizebox{0.7\hsize}{!}{\includegraphics{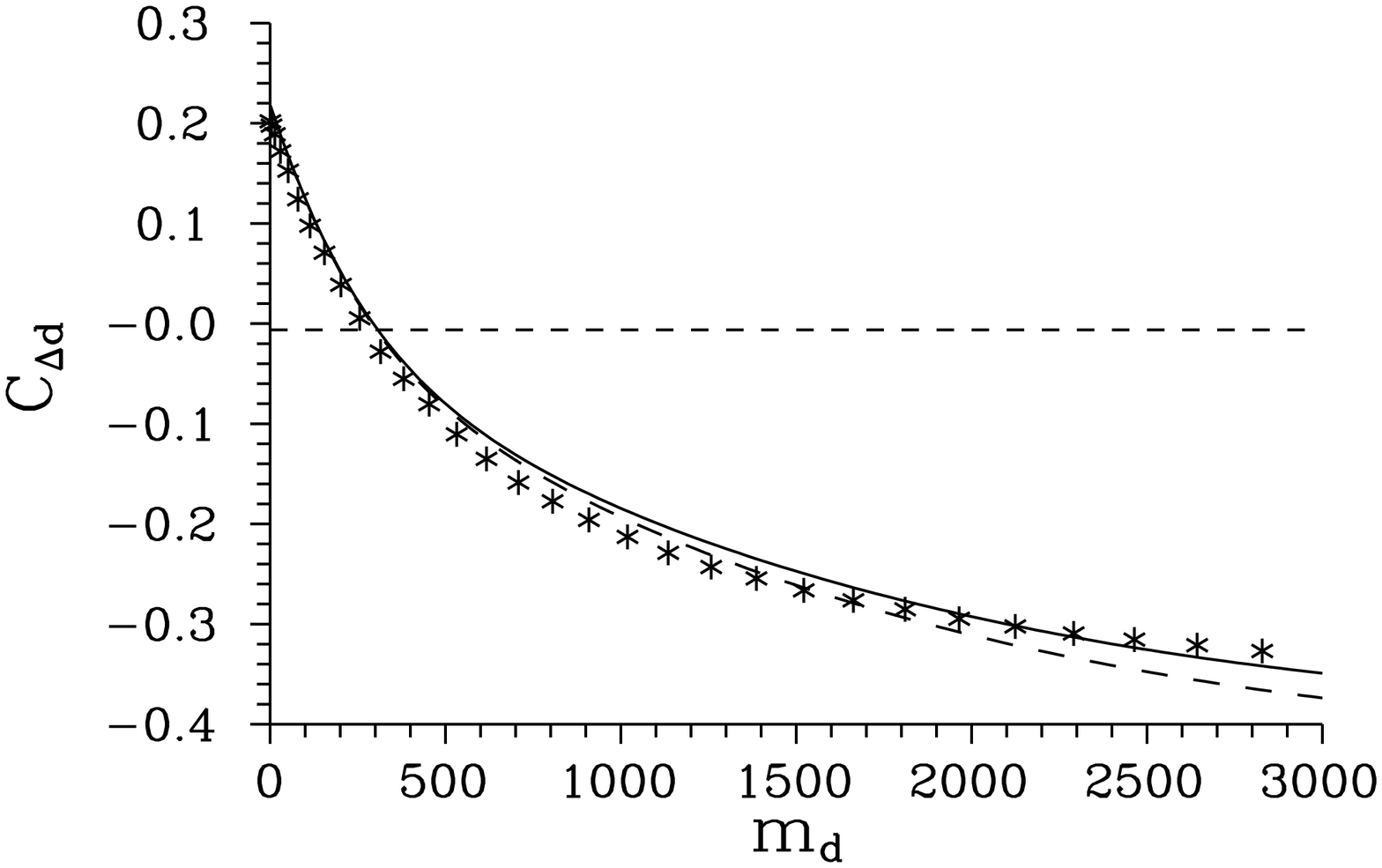}}}
  \vspace{2mm}
 \centerline{(b)\hspace{3mm}\resizebox{0.7\hsize}{!}{\includegraphics{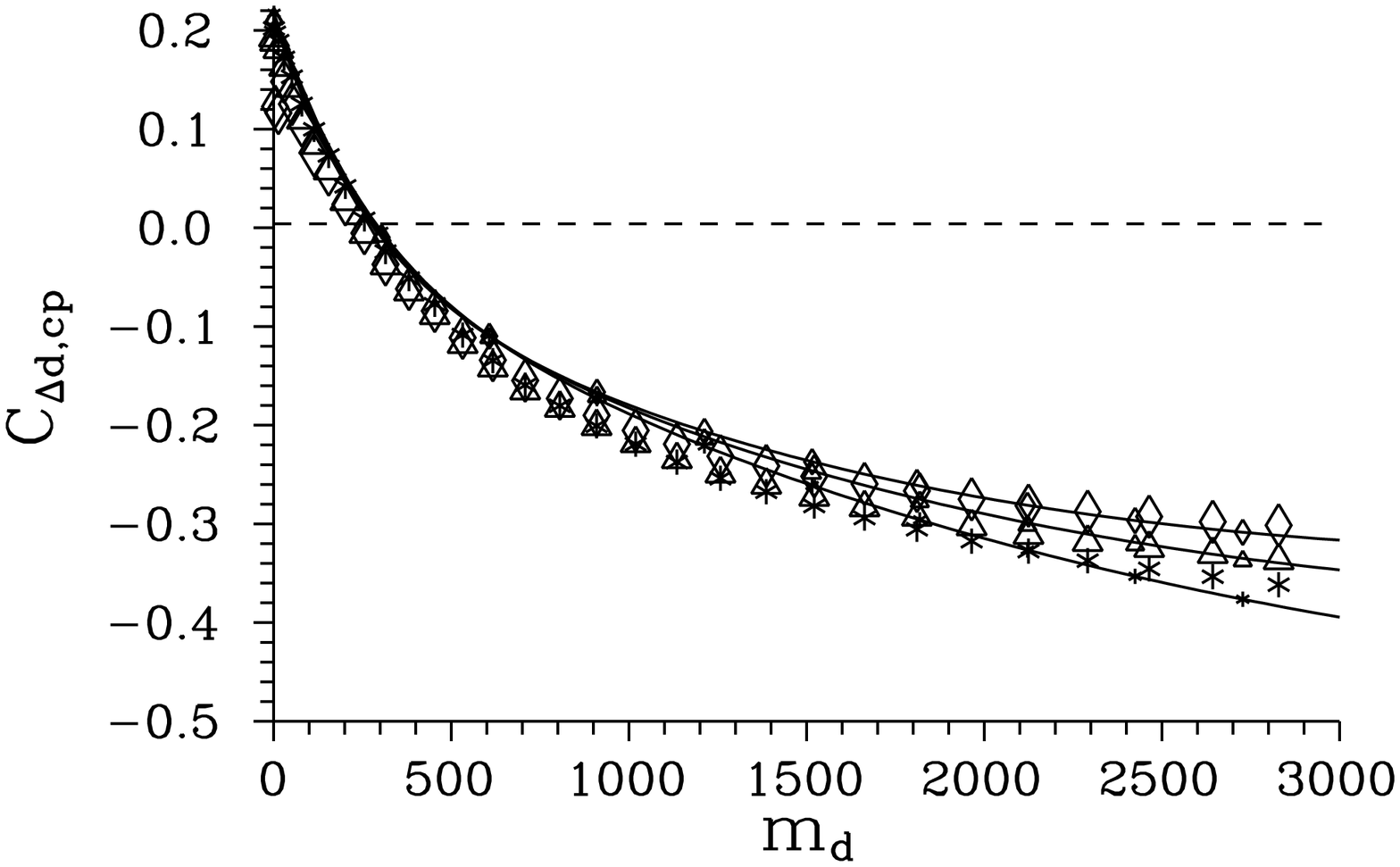}}}
  \vspace{2mm}
 \caption{(a) Covariance $ C_{\Delta d} $ of fluctuations of the numbers of unpaired signal and idler counts
  (relative experimental error is better than 2\%) and
  (b) covariances $ C_{\Delta d,c_{\rm p}} $ of fluctuations of the numbers of unpaired signal and idler counts
  conditioned by the detection of $ c_{\rm p} $ paired counts for $ c_{\rm p}
  = 0 $ ($ \ast $, relative experimental error is better than 2\%), $ c_{\rm p} = 1 $ ($ \triangle $, 4\%) and $ c_{\rm p} = 2 $
  ($ \diamond $, 6\%) as they depend of the number $ m_{\rm d} $ of detection pixels. Solid (dashed) curves originate in the
  refined (basic) model, isolated points characterize the experimental data. Dashed horizontal lines give the condition $ C=0 $.}
\label{fig6}
\end{figure}

More precise determination of the number $ m_{\rm c} $ of pixels
in the correlated area arises from the analysis of the joint
histograms of unpaired signal and idler counts. With the
increasing number $ m_{\rm d} $ of detection pixels the mean
number $ \langle d_{\rm s} \rangle $ of unpaired signal counts
decreases, whereas its relative variance $ \sigma_{d_{\rm s}} $
monotonically increases (see Fig.~\ref{fig5}).
However, as already discussed above, % in Sec.~III,
covariance $ C_{\Delta d} $ between the fluctuations of the
numbers of unpaired signal and idler counts changes its sign at
the edge of the correlated area, as documented in
Fig.~\ref{fig6}(a). The determination of number $ m_{\rm c} =
m_{\rm d}^0 $ of pixels in the correlated area can be considered
as a phenomenological definition of the extent of the correlated
area. This definition is robust in the sense that the change of
sign occurs nearly for the same values of $ m_{\rm d}^0(c_{\rm p})
$ also for the covariances $ C_{\Delta d,c_{\rm p}} $ determined
for the photocount distributions $ F_{\rm si}(d_{\rm s},d_{\rm
i};c_{\rm p},m_{\rm d}) $ conditioned by identification of $
c_{\rm p} $ paired counts in the frame [see Fig.~\ref{fig6}(b)].

\begin{figure}         % fig. 7
 \centerline{\resizebox{0.7\hsize}{!}{\includegraphics{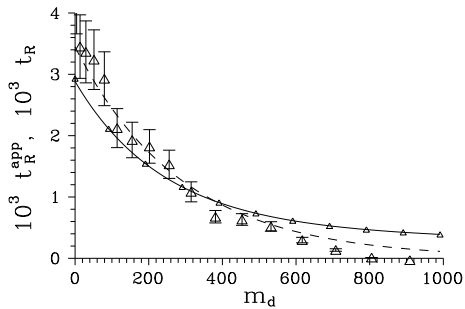}}}
 \vspace{2mm}
 \caption{Profile $ \tilde{t}_R^{\rm app} $ characterizing the correlated area given by Eq.~(\ref{37})
  as it depends on the number $ m_{\rm d} $ of detection pixels. Experimental points
  are plotted as isolated symbols ($ \triangle $), solid curve with $ \triangle $
  is derived for a theoretical Gaussian correlated area with $ m_{\rm c} = 250 $, for which
  $ m_{\rm d}^0 \approx 290 $. The corresponding ideal profile $ \tilde{t}_R = \exp(-m_{\rm d}/m_{\rm c})/m_{\rm c}  $
  is drawn by a dashed curve for comparison.}
\label{fig7}
\end{figure}

The obtained curve for the number $ \langle c_{\rm p} \rangle^{\rm
reg} $ of genuine paired counts allows us to reveal the profile $
\tilde{t}_R $ of the correlated area using Eq.~(\ref{37}) that
relies on the discrete numerical derivative of the experimental
dependence $ \langle c_{\rm p} \rangle^{\rm reg,exp}(m_{\rm d}) $.
As the theoretical curve $ \langle c_{\rm p} \rangle^{\rm reg,exp}
$ obtained for a two-dimensional Gaussian profile $ t $ is close
to the corresponding experimental dependence [see
Fig.~\ref{fig3}(a)] for $ m_{\rm d} \leq 2 m_{\rm d}^0 $, it is
more convenient to directly fit the experimental points with
function~(\ref{13}) to arrive at the number $ m_{\rm c} $ of
pixels in the correlated area. Both approaches are compared in
Fig.~\ref{fig7}.

\begin{figure}         % fig. 8
 \centerline{\resizebox{0.7\hsize}{!}{\includegraphics{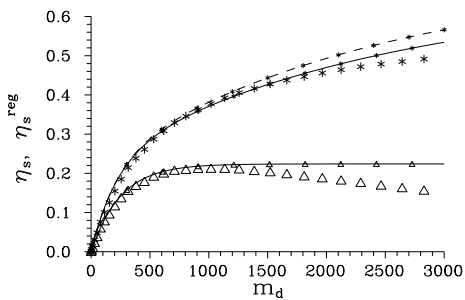}}}
 \vspace{2mm}
 \caption{Functions $ \eta_{\rm s} \equiv \langle c_{\rm p} \rangle / \langle c_{\rm i}
  \rangle $ ($ \ast $, relative experimental errors are better than 2\%)
  and $ \eta_{\rm s}^{\rm reg} \equiv \langle c_{\rm p} \rangle^{\rm reg} / \langle c_{\rm i}
  \rangle $ ($ \triangle $, 3\%) as they depend on the number $ m_{\rm d} $ of detection
  pixels. Solid (dashed) curves arise in the
  refined (basic) model, isolated points are derived from the experimental data.}
\label{fig8}
\end{figure}

Also detection efficiencies $ \eta_{\rm s} $ and $ \eta_{\rm i} $
appropriate for the signal and idler detection fields (strips),
respectively, can be deduced from the obtained experimental data.
In the scheme for absolute detector calibration suggested by
Klyshko \cite{Klyshko1980}, detection efficiency $ \eta_{\rm s} $
in the signal field is given as $ \langle c_{\rm p} \rangle /
\langle c_{\rm i} \rangle $ where $ \langle c_{\rm i} \rangle $
denotes the mean number of counts in the idler field (strip). As
the number $ \langle c_{\rm p} \rangle $ of paired counts
increases with the increasing number $ m_{\rm d} $ of detection
pixels, the curve $ \eta_{\rm s}(m_{\rm d}) $ also increases. In
the area around $ m_{\rm d}^0 $ it gives a good estimate $
\eta_{\rm s}^0 $ for the actual detection efficiency. According to
the curve $ \eta_{\rm s}(m_{\rm d}) $ plotted in Fig.~\ref{fig8},
we have $ \eta_{\rm s}^0 \approx 0.23 $. Provided that we
determine the efficiency $ \eta_{\rm s} $ using the number $
\langle c_{\rm p} \rangle^{\rm reg} $ of genuine paired counts,
the appropriate value is ideally revealed for $ m_{\rm d} = N $
(see Fig.~\ref{fig8}). However, in our realistic case with
underestimated numbers $ \langle c_{\rm p} \rangle^{\rm reg,exp} $
of genuine paired counts, we estimate the detection efficiency $
\eta_{\rm s} $ by the maximum of function $ \eta_{\rm s}(m_{\rm
d}) $ reached for the numbers $ m_{\rm d} $ of detection pixels
slightly larger than the number $ m_{\rm c} $ of pixels in the
correlated area. We note that the analysis also gives $ \eta_{\rm
i}^0 \approx 0.23 $ for the idler field.

\begin{figure}         % figs. 9a,b,c,d
 \centerline{\resizebox{0.46\hsize}{!}{\includegraphics{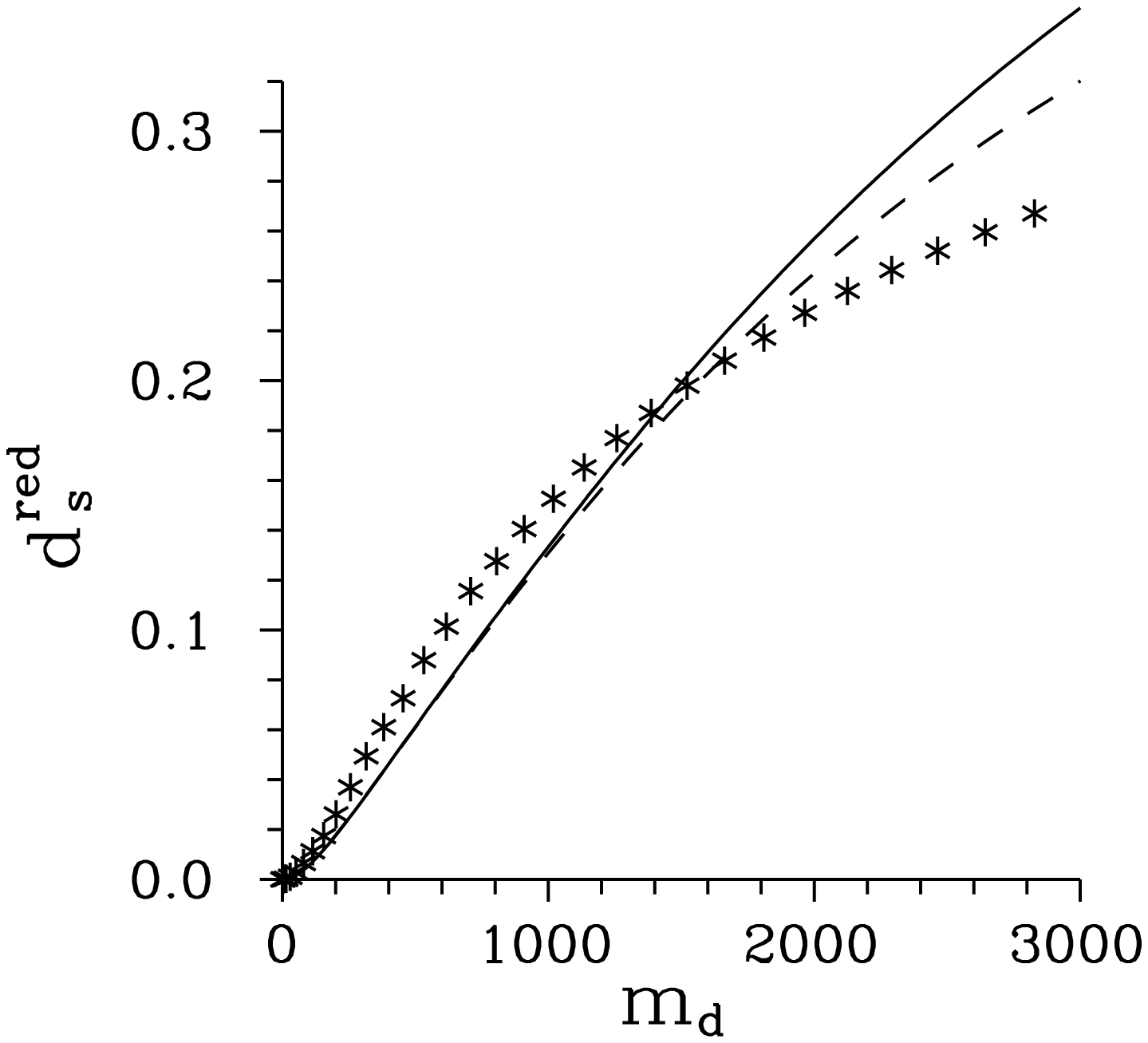}}
 \hspace{3mm}
 \resizebox{0.46\hsize}{!}{\includegraphics{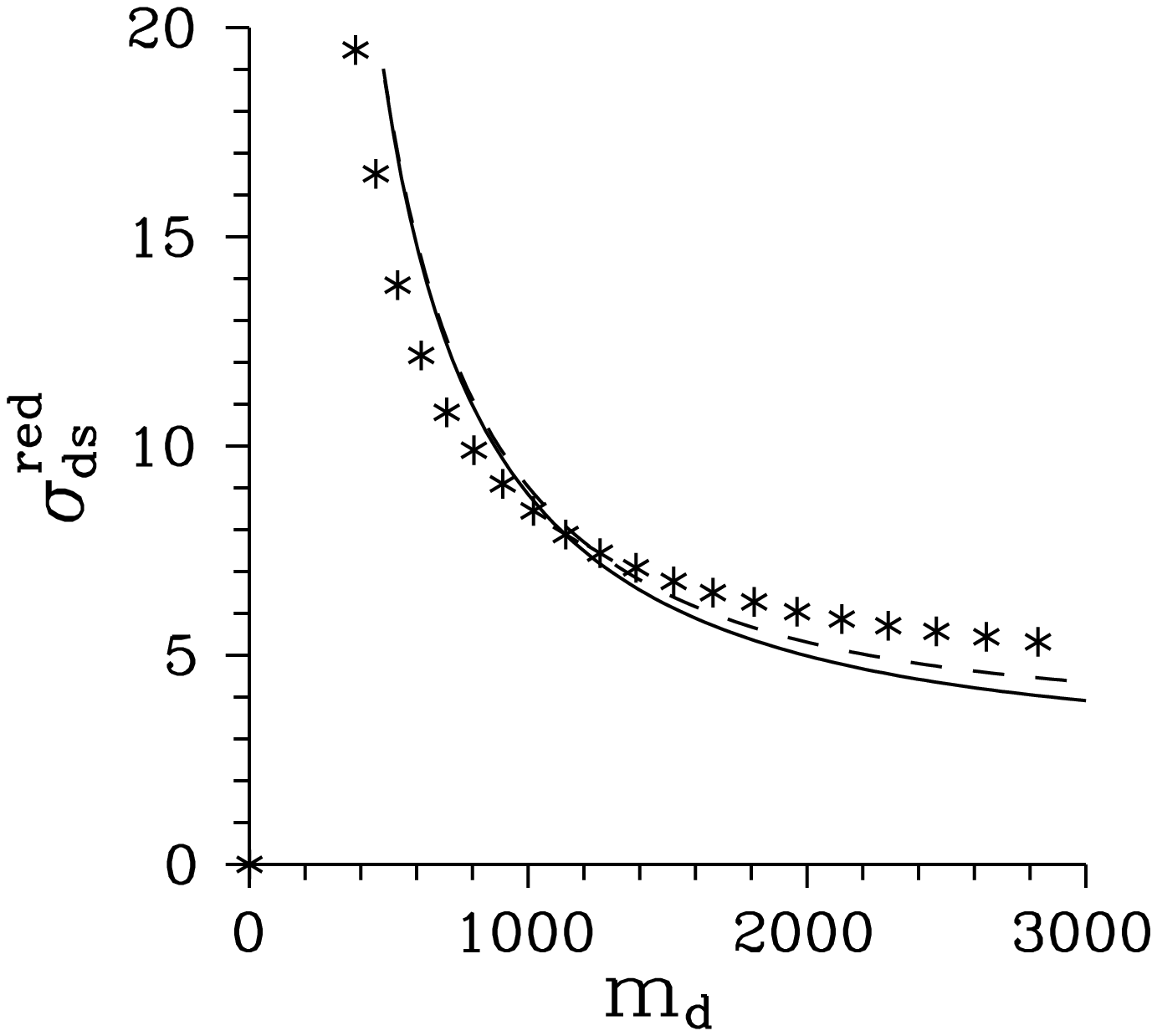}}}
 \centerline{(a) \hspace{.2\textwidth} (b)}
 \vspace{2mm}
 \centerline{\resizebox{0.46\hsize}{!}{\includegraphics{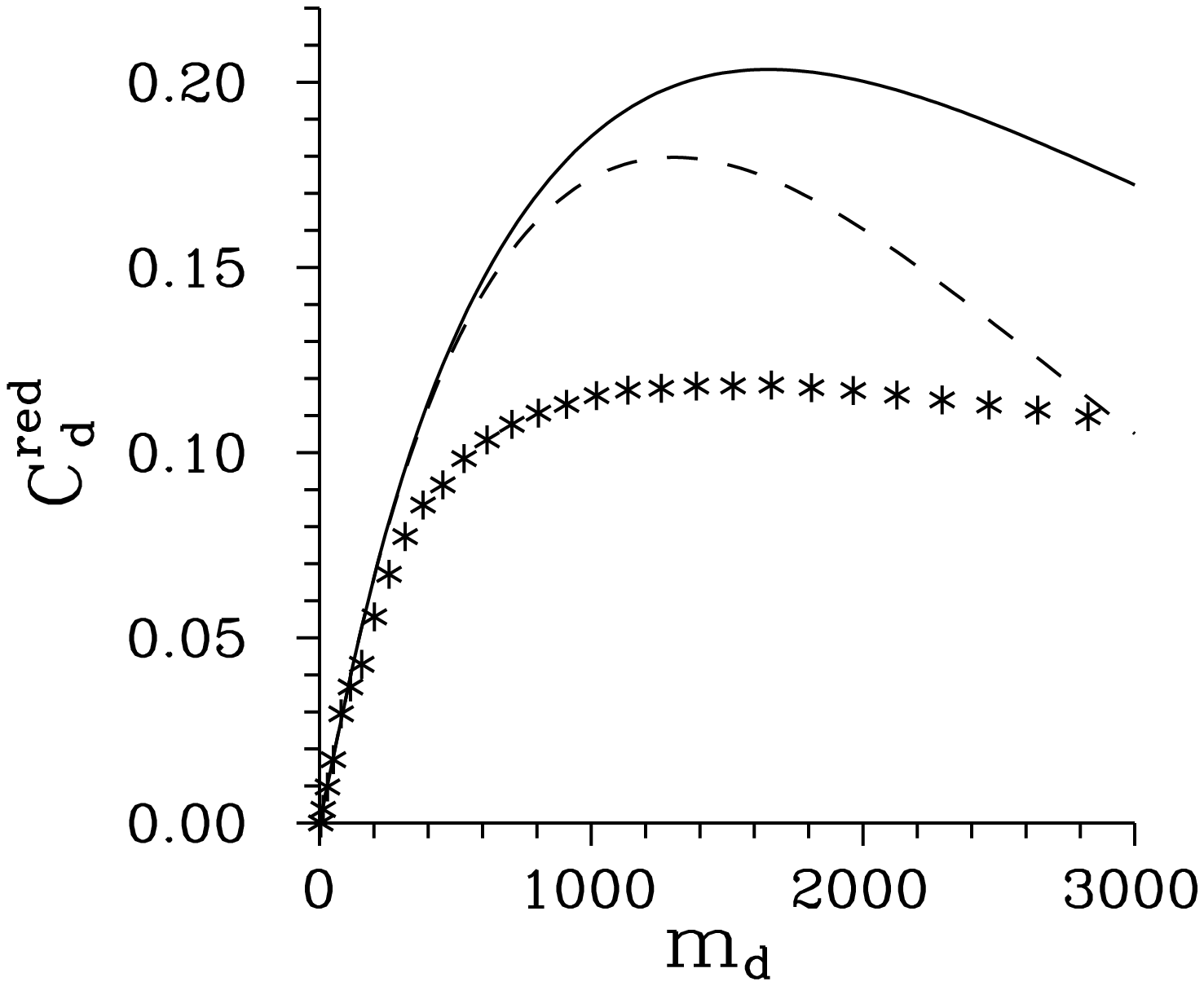}}
 \hspace{3mm}
 \resizebox{0.46\hsize}{!}{\includegraphics{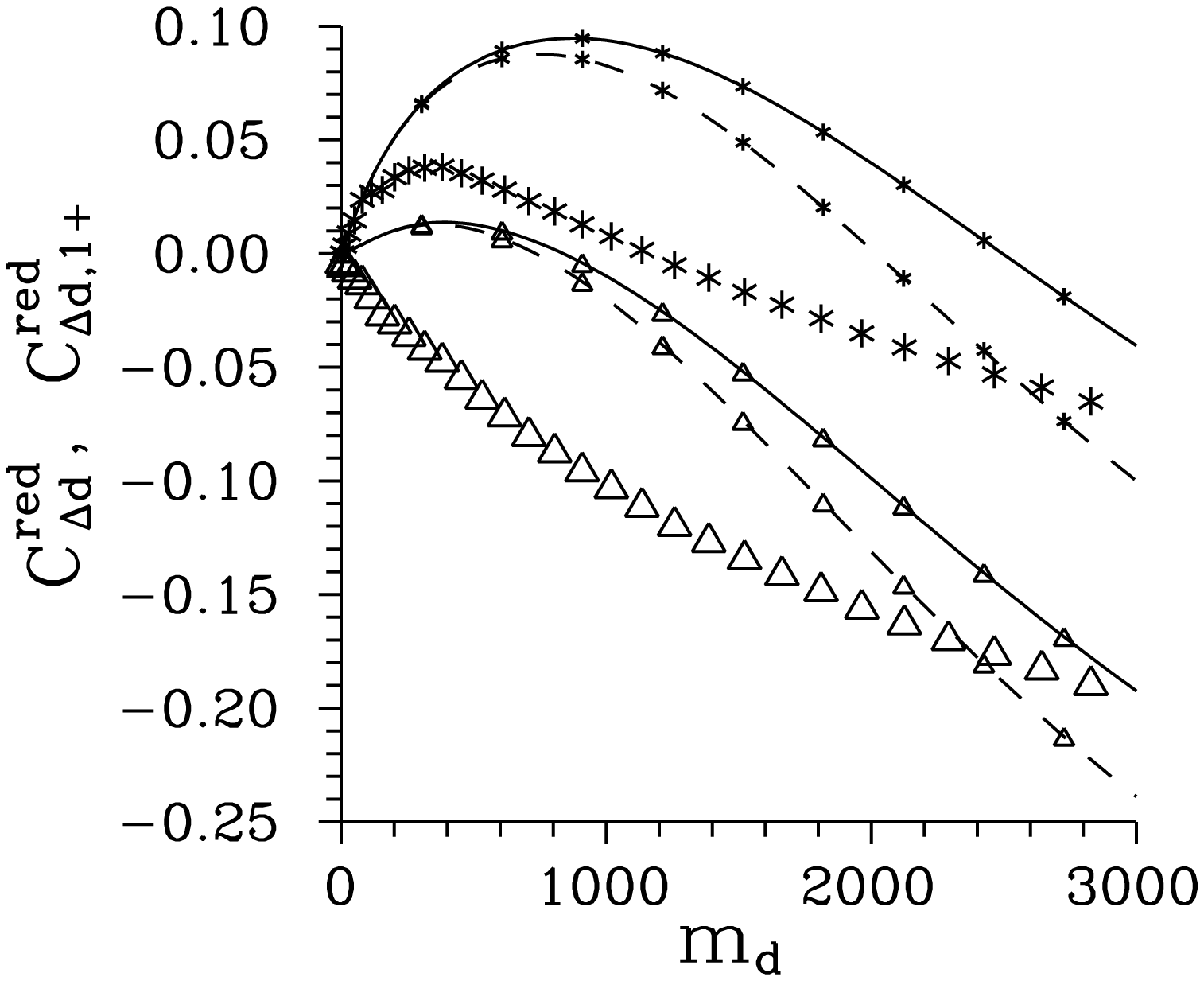}}}
 \centerline{(c) \hspace{.2\textwidth} (d)}
 \vspace{2mm}
 \caption{(a) Mean number $ \langle d_{\rm s} \rangle^{\rm red} $ of unpaired signal counts
 [relative experimental errors are lower than 4\%],
 (b) relative variance $ \sigma_{d_{\rm s}}^{\rm red} $ of their distribution [4\%],
 (c) covariance $ C_d^{\rm red} \equiv \langle d_{\rm s}d_{\rm i}\rangle / \sqrt{\langle d_{\rm s}^2 \rangle
  \langle d_{\rm i}^2\rangle } $ of the numbers of unpaired signal and idler counts ([4\%]), and (d)
  covariance $ C_{\Delta d}^{\rm red} $ of fluctuations of the numbers of unpaired signal
  and idler counts according to Eq.~(\ref{35}) ($ \ast $, [8\%]) and
  covariance $ C_{\Delta d,1+}^{\rm red} $ of fluctuations of the numbers of unpaired signal and idler
  counts in frames with at least one detected paired count ($ \triangle $, [10\%])
  as functions of the number $ m_{\rm d} $ of detection pixels assuming the noise
  reduction given in Eqs.~(\ref{30})---(\ref{32}).
  Solid (dashed) curves originate in the refined (basic) model, isolated points are obtained from the experimental data.}
\label{fig9}
\end{figure}

The numbers $ \langle d_{\rm s} \rangle $ and $ \langle d_{\rm i}
\rangle $ of unpaired counts in the signal and idler fields
(strips), respectively, represent from the point of view of
photon-pair detection unwanted noise. However, these numbers can
conveniently and substantially be reduced when we consider them
only inside the detection areas drawn around the paired counts
[see Eqs.~(\ref{30})---(\ref{32})], as it clearly follows from the
comparison of curves in Figs.~\ref{fig5}(a) and \ref{fig9}(a).
According to the experimental points plotted in
Fig.~\ref{fig9}(a), decrease of the detection area from 3000
pixels to one third reduces the number $ \langle d_{\rm s}
\rangle^{\rm red} $ of noise signal counts roughly to one half.
For numbers $ m_{\rm d} $ of detection pixels comparable to the
number $ m_{\rm c} $ of pixels inside the correlated area,
additional reduction of the noise by one half is observed. This
noise reduction can be exploited when determining the distribution
of photon pairs from the experimental joint signal-idler
photocount histograms once the extent of the correlated area ($
m_{\rm c} $) is known (or estimated). However, this procedure has
to be done with care as, for the detection area smaller than the
correlated area, the noise reduction is accompanied by a
relatively large increase of photocount fluctuations [see
Fig.~\ref{fig9}(b)]. The noise reduction qualitatively changes the
behavior of cross-correlations between the unpaired signal and
idler counts for small detection areas. Both covariances $
C_d^{\rm red} $ of the numbers of unpaired signal and idler counts
and $ C_{\Delta d}^{\rm red} $ of fluctuations of the numbers of
unpaired signal and idler counts tend to zero for small numbers $
m_{\rm d} $ of detection pixels, as documented in
Figs.~\ref{fig9}(c) and \ref{fig9}(d). On the other hand,
relatively weak correlations are observed in the whole range of
numbers $ m_{\rm d} $ of detection pixels. Mutual comparison of
the experimental points and two theoretical curves indicates that
the covariance $ C_d^{\rm red} $ of the numbers of unpaired signal
and idler counts is quite sensitive to the experimental conditions
as well as the detailed structure of the theoretical model.
Detailed geometry of the detection fields (strips), that is not
treated in the model, probably plays an important role in the
explanation of large deviations among the values of covariance $
C_d^{\rm red} $ plotted in Fig.~\ref{fig9}(c). Deviations among
the values of covariances $ C_{\Delta d}^{\rm red} $ and $
C_{\Delta d,1+}^{\rm red} $ of fluctuations of the numbers of
unpaired signal and idler counts are even larger (for the
definition of $ C_{\Delta d,1+}^{\rm red} $, see the caption to
Fig.~\ref{fig9}). They have their origin in larger discrepancies
between the experimental and theoretical mean numbers $ \langle
d_{\rm s}\rangle^{\rm red} $ [see Fig.~\ref{fig9}(a)] and $
\langle d_{\rm i}\rangle^{\rm red} $ and signal-idler correlation
function $ \langle d_{\rm s}d_{\rm i}\rangle^{\rm red} $ [compare
Fig.~\ref{fig9}(c)] that are determined for unpaired counts. Only
qualitative agreement can be seen in the graph of
Fig.~\ref{fig9}(d) when the experimental points are compared with
the theoretical predictions. However, neither the experimental
points nor the theoretical curves indicate the ability of
covariances $ C_{d}^{\rm red} $, $ C_{\Delta d}^{\rm red} $ and $
C_{\Delta d,1+}^{\rm red} $ to provide the extent of the
correlated area. Proper description of the observed dependencies
of these covariances lies beyond the developed model. On the other
hand, the absence of reliable description of weak correlations
between the unpaired signal and idler photocounts does not
seriously restrict the application of the noise reduction in
photon-pair counting.

\begin{figure}         % fig. 10
 \centerline{\resizebox{0.7\hsize}{!}{\includegraphics{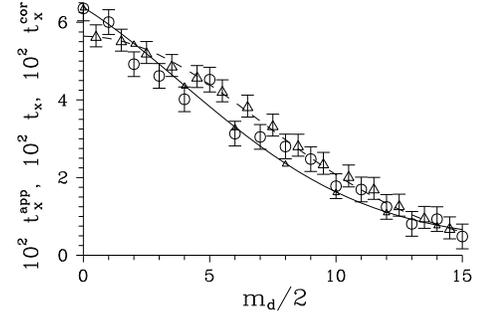}}}
 \vspace{2mm}
 \caption{Profile $ t_x^{\rm app} $ of the correlated area along the $ x $ axis
  determined by Eq.~(\ref{38}) as it depends on the number $ m_{\rm d} $ of detection pixels. Experimental points
  are plotted as isolated symbols ($ \triangle $), solid curve with $ \triangle $
  is derived for the theoretical Gaussian correlated area with $ m_{\rm c}/2 = 10 $ ($ m_{\rm d}^0/2 = 10 $).
  The corresponding ideal profile $ t_x = 2/(\sqrt{\pi}m_{\rm c})\exp(-m_{\rm d}^2/m_{\rm c}^2) $
  is drawn by a dashed curve. For comparison, experimental points of $ t_x^{\rm cor} $ obtained from direct evaluation of the
  intensity cross-correlation function are shown ($ \circ $).}
\label{fig10}
\end{figure}

The developed method for revealing the profiles of correlated
areas can easily be modified to provide cuts through
two-dimensional intensity spatial cross-correlation functions. In
this case rectangular detection areas with the varying number of
pixels in one direction and fixed number of pixels in the
perpendicular direction (ideally just one) is applied. For
example, rectangular detection areas extending over $ m_{\rm d} $
detection pixels along the $ x $ axis and covering just one pixel
along the $ y $ axis give us the profile of the intensity
cross-correlation function along the $ x $ axis. The experimental
profile $ t_{x}^{\rm app} $ obtained according to Eq.~(\ref{38})
is compared in Fig.~\ref{fig10} with the profile $ t_{x}^{\rm cor}
$ arising in direct evaluation of the intensity spatial
cross-correlation function that relies on the subtraction of a
constant background (for details, see
\cite{Haderka2005,Hamar2010}). Also theoretical curves
characterizing a suitable Gaussian profile are drawn in
Fig.~\ref{fig10} for comparison. Compared to the above analyzed
two-dimensional case with circular detection areas, we have by two
orders in magnitude lower numbers of genuine photon pairs per
one-dimensional 'frame'. On the other hand, the number of such
'frames' contained in the experimental ensemble is by two orders
in magnitude greater than the number of the original
two-dimensional frames. This fact gives better experimental
precision. Moreover, if the applied rectangular detection area
extends over several pixels in the perpendicular direction, we
arrive at averaged (smoothed) profiles of intensity
cross-correlation functions. Also, larger numbers of genuine
photon pairs are met in this case. We note that the method can be
applied separately to the left- and right-hand sides of the
correlated area (from its center) to reveal its profile
completely.

The developed method for revealing profiles of intensity
cross-correlation functions through the determination of the
numbers of genuine paired photocounts requires in principle much
lower amount of experimental data compared to the usual approach
that is based on accumulating random accidental paired photocounts
until they form a constant background
\cite{Haderka2005,Hamar2010}. This advantage is achieved by a
sophisticated processing of the experimental data that estimates
mean numbers of accidental paired photocounts.

The comparison of the remaining experimental and theoretical
quantities independent on the number $ m_{\rm d} $ of detection
pixels reveals, together with the curves already presented in the
graphs, good agreement between both models and the measured
quantities (mean number of all signal counts: $ \langle c_{\rm
s}\rangle^{\rm exp} = 2.00\pm 0.02 $, $ \langle c_{\rm s}\rangle =
2.01 $; relative variance of all signal counts: $ \sigma_{c_{\rm
s}}^{\rm exp} = 0.51 \pm 0.01 $, $ \sigma_{c_{\rm s}} = 0.51 $;
covariance of fluctuations of the numbers of all signal and idler
counts: $ C^{\rm exp} = 0.200 \pm 0.004 $, $ C = 0.22 $; noise
reduction factor $ R^{\rm exp} = 0.81 \pm 0.02 $, $ R = 0.80 $).
The overall comparison of the experimental points with the curves
determined by both models leads to the general conclusion that the
experimental points are fitted better by the curves of the refined
model. Both models allow for reliable quantification of the
processes of detection of correlated fields with
photon-number-resolving detectors endowed with spatial resolution.

\section{Conclusions}

We have developed a suitable statistical model (in two variants)
that allows for quantifying the role of spatial correlations in
the observed joint signal-idler photocount distributions of a weak
twin beam. In the model the detected counts are divided into those
corresponding to genuine paired photocounts, accidental paired
photocounts and unpaired photocounts. The model allows to separate
the genuine paired photocounts from the remaining ones and
subsequently to recover the profile of intensity spatial
cross-correlation functions. The determination of intensity
cross-correlation functions has been demonstrated in one- and
two-dimensional geometries. In parallel with the quantification of
spatial correlations, the principle of reduction of the noise
occurring in photon-number-resolving detection with the help of
spatial correlations has been experimentally demonstrated.

\section{Acknowledgments}

The authors thank M. Hamar for his help with the experiment. The
authors acknowledge projects P205/12/0382 and 15-08971S of
GA\v{C}R and LO1305 of M\v{S}MT \v{C}R for support.

% \bibliography{perina}

%merlin.mbs apsrev4-1.bst 2010-07-25 4.21a (PWD, AO, DPC) hacked
%Control: key (0)
%Control: author (0) dotless jnrlst
%Control: editor formatted (1) identically to author
%Control: production of article title (0) allowed
%Control: page (1) range
%Control: year (0) verbatim
%Control: production of eprint (0) enabled
%

\end{document}